\documentclass[10pt]{iopart}

\usepackage{geometry}

\usepackage{graphicx}
\usepackage{amsmath,amssymb,bm}
\usepackage{mathrsfs}
\usepackage[hidelinks]{hyperref}
\usepackage{cite}
\usepackage[title]{appendix}
\usepackage{comment}

\usepackage{color}
\usepackage[dvipsnames]{xcolor} 
\usepackage{outlines} 

\newcommand{\blue}[1]{\textcolor{black}{#1}}
\newcommand{\red}[1]{\textcolor{black}{#1}}

\begin{document}

\title[Simulating X-point radiator turbulence]{Simulating X-point radiator turbulence}

\author{K. Eder$^*$, W. Zholobenko, A. Stegmeir, M. Bernert, D. Coster, F. Jenko, the ASDEX Upgrade Team$^{1)}$, and the EUROfusion Tokamak Exploitation Team$^{2)}$} 

\address{Max-Planck-Institut für Plasmaphysik, Boltzmannstraße 2, 85748 Garching, Germany

$^{1)}$ see author list of H. Zohm \textit{et al.}, 2024 \textit{Nucl. Fusion} \textbf{64} 112001, \url{https://doi.org/10.1088/1741-4326/ad249d}

$^{2)}$ see author list of E. Joffrin \textit{et al.}, 2024 \textit{Nucl. Fusion} \textbf{64} 112019, \url{https://doi.org/10.1088/1741-4326/ad2be4}

$^*$ Corresponding author: konrad.eder@ipp.mpg.de}

\date{\today}

\begin{abstract}

Coupling a high-performance burning plasma core to a detached boundary solution is critical for realizing magnetic confinement fusion power. 
Predictive simulations of the edge and scrape-off layer are therefore essential and must self-consistently account for turbulence and the interplay between the plasma, neutral gas, and impurities. 
We present results on controlled full detachment in ASDEX Upgrade with an X-point radiator (XPR), obtained with the edge turbulence code GRILLIX.
\red{Assuming a fixed nitrogen concentration (in terms of the electron density) in coronal equilibrium, two simulations are discussed: they exhibit dense nitrogen radiation fronts, located $5$ and $12\,\mathrm{cm}$ above the X-point, accounting for $80\%$ of the input heating power.} 
In validations against density, temperature, and bolometry measurements, the simulations show good agreement and reproduce the detached divertor conditions observed in the experiment.
Neutral gas is critical for achieving detachment and modulating the height of the XPR front, in agreement with previous SOLPS-ITER transport modeling and analytical power balance studies. 
In addition, the front structure is highly dynamic due to turbulence, consisting of ionizing and radiative mantles surrounding intermittent cold spots of recombining plasma.
Near the detachment front, density and temperature fluctuation amplitudes exceed the background by more than 400\%, compared to 40\% in an attached reference case.
\red{The radial electric field shifts inward, poloidal symmetry of the electrostatic potential is broken (inducing strong radial flows around the XPR), and radial particle and heat transport into the low-field side scrape-off layer increases. These effects may explain the ELM suppression observed in the H-mode XPR regime.}

    
\end{abstract}

\noindent To be submitted to: \textit{Nuclear Fusion}

\noindent Last edited: \today

\maketitle


\section{Introduction}
\label{section:introduction}

A key step on the path to fusion power is the integration of a stationary burning plasma core with a power exhaust solution \cite{viezzer2023}. 
Projections to reactor-scale machines \cite{zohm2013, pitts2019} suggest that particle and heat loads onto plasma-facing components will have to be mitigated \cite{loarte2007, wischmeier2015}.
Divertor detachment \cite{stangeby2000}, along with advanced divertor concepts \cite{kuang2020, zohm2021}, therefore remain vital components of future exhaust scenarios. 
Numerical simulations --- essential for interpreting experiments, guiding plasma operation, and extrapolating to new devices --- thus need to resolve detached conditions, involving atomic processes such as impurity radiation and plasma-neutral interactions \cite{krasheninnikov2017, pshenov2017}. 
Contemporary transport codes, such as SOLPS-ITER \cite{wiesen2015}, SOLEDGE2D-EIRENE \cite{bufferand2013}, or EMC3-EIRENE \cite{feng2014} have risen to the task and are accordingly well-integrated into scenario design on current machines \cite{bucalossi2022, stroth2022scenario}. However, they still rely on heuristic transport coefficients to recover experimental profiles. 
To perform \textit{predictive} power exhaust studies, it becomes necessary to resolve the turbulent dynamics in the edge and scrape-off layer (SOL) that underpin much of the radial transport. 
This involves a dramatic increase in simulation complexity and computational cost compared to transport simulations, made more convoluted when coupled to atomic and molecular models.
Nonetheless, global turbulence codes have since begun to implement neutral gas and impurity physics \cite{wersal2015, kvist2024, quadri2024, eder2025}, and pioneering studies on detached plasma turbulence are already underway \cite{mancini2023, quadri2024}. 

This manuscript and the turbulence simulations discussed herein continue our joint effort to understand turbulence in detached conditions, with a special focus on \textit{X-point radiation}.
Strong impurity radiation near or above the X-point (radiating a majority of input heating power) has long been observed experimentally in detached conditions across multiple tokamaks, with both carbon \cite{kallenbach1995, lowry1997, pitts1999, park2023} and metal \cite{goetz1996, reimold2015a, gloeggler2019, bucalossi2022} walls.

Only recently, however, it was discovered that the radiative structure can be fully stabilized by feedback control \cite{reimold2015a, bernert2021}, differentiating it from the unstable radiative phenomena known as MARFEs \cite{lipschultz1984}.
Consequently, the X-point radiator (XPR) \cite{bernert2025} is nowadays regarded as a distinct operational scenario instead of merely a phenomenon in detached discharges.
Feedback-controlled stable X-point radiation has now been achieved in various tokamaks \cite{bernert2021, bernert2023, rivals2024,  bosman2024, reimerdes2024} and successfully reproduced in transport simulations with SOLPS-ITER \cite{senichenkov2021a, pan2022, sun2023, poletaeva2024} and SOLEDGE3X \cite{rivals2024}. 
Similar radiating structures have been observed experimentally at X- and O-points in island-diverted stellarator configurations of Wendelstein 7-X and the Large Helical Device and are reproducible with EMC3-EIRENE \cite{feng2024, winters2024, kobayashi2013}.
The detached XPR state can be sustained throughout the L-H transition \cite{bernert2025}, and was found to suppress edge localized modes (ELMs) in H-mode conditions if the radiating structure exceeds a certain height ($\approx 7\,\mathrm{cm}$ above the X-point in ASDEX Upgrade) \cite{bernert2021}. 
This opens the door for entirely Type-I ELM-free ramp-ups from L-mode into the H-mode XPR conditions, or into other high-performance no-ELM/small-ELM modes, such as the Quasi-Continuous Exhaust (QCE) \cite{faitsch2021}, or enhanced $D_\alpha$ H-mode \cite{greenwald1999} regimes.
Alternatively, by capitalizing on the inherent power exhaust properties of the XPR, main X-points can be placed close to the wall (known as a Compact Radiative Divertor \cite{lunt2023}), resulting in increased plasma volume and, thus, performance.

In view of these exciting developments, it is imperative that turbulence codes keep pace and help illuminate the complex dynamics at play.
To this end, we present first-of-its-kind turbulence simulations in detached X-point radiating conditions, performed with the edge turbulence code GRILLIX \cite{stegmeir2019}.
The code itself will be introduced in Section \ref{section:model}. There, we also document code developments vital for simulating detachment, namely the introduction of impurity radiation in addition to the previously implemented neutral gas model, along with numerical improvements.
The remainder of this manuscript is then structured as follows: 
Section \ref{section:application} first describes the simulation setup, including parameter settings, identification of the XPR state, and modulation of XPR height. 
In Section \ref{section:validation}, the simulations are validated against experimental profiles and bolometry data.
\red{Section \ref{section:analysis} discusses additional comparative analyses of fluctuation amplitudes, radial transport, and electrostatic potential across simulations of varying XPR states.}
Finally, we conclude our findings in Section \ref{section:conclusion} and provide a look at the work ahead.


\section{The GRILLIX edge turbulence code}
\label{section:model}

GRILLIX \cite{stegmeir2019} is a global full-\textit{f} drift-fluid turbulence code applied in the edge and scrape-off layer (SOL). 
It is based on the Flux-Coordinate-Independent (FCI) approach \cite{stegmeir2016}, which is computationally beneficial for resolving anisotropic 3D plasma turbulence \cite{stegmeir2023} and compatible with both complex diverted tokamak \cite{body2019} and stellarator \cite{stegmeir2025} geometries. 
In the FCI approach, the grid resolution is unaffected by the field singularity at the X-point, making the study of XPR turbulence a particularly fitting objective. 
GRILLIX consists of an electromagnetic drift-fluid plasma model \cite{zhang2024, zholobenko2024} coupled to a fluid neutrals model \cite{zholobenko2021b, eder2025}. 
For conciseness, we omit the full system of GRILLIX equations and instead refer to \cite{zholobenko2024} and \cite{eder2025} for the plasma and neutrals models, respectively. 
The former is based on the global, electromagnetic, drift-reduced Braginskii equations \cite{braginskii1965,zeiler1997}, with its closure extended to low collisionality conditions by including a neoclassical ion viscosity and limiting the parallel heat flux to a free-streaming fraction \cite{zholobenko2024}.
Notice that the XPR regime is highly collisional \cite{bernert2023}, for which the fluid approach is valid by construction. 
Moreover, strong fluctuation levels with strong gradients are expected in detached conditions (see Section \ref{section:analysis}), such that a full-\textit{f} description as provided by GRILLIX is mandatory.

Regarding the physics of neutral gas, GRILLIX implements a 3-moment fluid approach \cite{horsten2017, uytven2020} that accounts for plasma-neutrals interactions through charge-exchange, ionization, and recombination. The resulting sources/sinks of particles, momentum, and heat induce a wide range of effects, including the alteration of background plasma profiles \cite{zholobenko2021b, stotler2017} and the damping of turbulent zonal flows \cite{zhang2020}.
Cold neutrals from the divertor further play important roles in reaching detachment by charge-exchange cooling and momentum dissipation \cite{eder2025}.
Facilitating detachment simulations in GRILLIX, however, required additional steps beyond the implementation outlined so far. These will be discussed in the following paragraphs.

\subsection{Impurity radiation}
\label{section:model:impurity}

In simplified terms, an XPR is understood to result from the balance of power conducted into the XPR volume and power radiated out of the volume by impurities \cite{stroth2022xpr, pan2022}. 
Analytical models \cite{stroth2022xpr, stroth2025} suggest that a stable power balance can already be recovered by assuming impurities to represent a constant fraction of plasma density, $n_\mathrm{imp} = c_\mathrm{imp} n$. The impurity-radiated power density then reads
\begin{equation}\label{eq:impurity_radiation}
    p^\mathrm{imp}_\mathrm{rad} = - L_\mathrm{imp}\left(T_\mathrm{e}\right) n^2 c_\mathrm{imp} \, ,
\end{equation}
where the effective radiation rate coefficient $L_\mathrm{imp}$ is derived for nitrogen in coronal equilibrium \cite{adas}\footnote{While this work focuses on nitrogen-seeded plasmas, XPRs have also been observed experimentally by seeding Neon, Argon, or Krypton \cite{bernert2023}.}. It depends strongly on electron temperature $T_\mathrm{e}$, peaking at $\approx 7 - 15\,\mathrm{eV}$, with a steep drop in either direction. 

For turbulence simulations shown herein, the time evolution of electron temperature in GRILLIX has been modified to include impurity radiation loss according to \eqref{eq:impurity_radiation}. 
A self-consistent description of impurities based on the Zhdanov fluid closure is envisioned as future work \cite[Chapter~5.2]{makarov2024}. 
Note that the simulation domain does not extend into the plasma core (see Section \ref{section:application:setup}), where the assumption of constant $c_\mathrm{imp}$ does not apply. 
Additionally, as will be shown in Section \ref{section:application}, radiation beyond the XPR structure is marginal due to the steep reduction of $L_\mathrm{imp}$ with typical confinement temperatures $T_\mathrm{e} > 20 \, \mathrm{eV}$. 

\subsection{Numerical stability}

In detachment, electron temperature drops to $T_\mathrm{e} < 3\,\mathrm{eV}$ while the confined plasma remains at hundreds of eV.
At such low temperatures, the parallel resistivity in Ohm's law becomes stiff if treated explicitly, since it scales inversely with electron temperature, $\eta_\parallel \propto T_\mathrm{e}^{-3/2}$. In terms of physics, the limitation arises due to resistive perpendicular diffusion of the perturbed magnetic field, i.e.~due to over-damped shear-Alfvén waves. To avoid restrictions on the time-step size, we now treat the term with an implicit scheme, outlined in the appendix \ref{parallel_resistivity}.
The semi-explicit treatment of pressure diffusion in the neutrals model proved another hurdle, as it becomes stiff in the presence of large temperature gradients. This has also been resolved by an implicit scheme, as documented in \cite{eder2025}.

\subsection{Lower limit on neutrals diffusivity}

The GRILLIX neutrals model \red{\cite{eder2025}} describes the time evolution of neutrals density $N$, parallel momentum $\Gamma_\mathrm{N} \mathbf{b}$, and pressure $p_\mathrm{N}$.
Neutrals diffuse perpendicular to the magnetic field $\mathbf{B} = \mathbf{b} B$ due to charge-exchange with ions, which can be described in a generalized form for any neutrals moment $u_\mathrm{N} \in \left\{ N, \Gamma_\mathrm{N}, p_\mathrm{N} \right\}$ as
\begin{equation}\label{eq:diffusion}
    \frac{\partial u_\mathrm{N}}{\partial t} = \nabla \cdot D_{u_\mathrm{N}} \nabla_\perp \left( u_\mathrm{N} T_\mathrm{N} \right) + \cdots \, ,
\end{equation}
\red{dependent on neutrals temperature $T_\mathrm{N}$ and charge-exchange diffusion coefficient $D_\mathrm{u_N} \propto \left( n \langle \sigma v \rangle_\mathrm{cx} \right)^{-1}$ as defined in \cite{eder2025}.}
To mimic kinetic non-local spreading that is not intrinsically captured in the local fluid approach, we apply a lower limit on diffusivity by introducing a conditional amplification factor,
\begin{equation}\label{eq:diffusion_limiter}
    \nabla \cdot \frac{\max\left( T_\mathrm{N} , T_\mathrm{N, min}\right)}{T_\mathrm{N}}D_{u_\mathrm{N}} \nabla_\perp \left( u_\mathrm{N} T_\mathrm{N} \right) \, ,
\end{equation}
dependent on the neutrals temperature threshold parameter $T_\mathrm{N, min}$. 





\section{Simulations}
\label{section:application}

\subsection{Setup}
\label{section:application:setup}

Simulations shown in this work are based on ASDEX Upgrade discharge \#40333, performed in favorable lower single null configuration ($\nabla B$ drift pointing towards the X-point), with plasma current $I_p = 0.8 \, \mathrm{MA}$, toroidal magnetic field on axis $B_\mathrm{tor} = 2.4 \, \mathrm{T}$ and edge safety factor $q_{95} = 4.63$. 
The discharge features a fully detached L-H transition with substantial X-point radiation before, throughout, and after the transition \cite{bernert2025}. We simulate the L-mode phase, with the magnetic equilibrium reconstructed from $2.4 \, \mathrm{s}$. At that time, the power input equals $1.7 \, \mathrm{MW}$ of ohmic heating and electron cyclotron resonance heating. $0.4 \, \mathrm{MW}$ of power is radiated in the core at normalized poloidal flux $\rho_\mathrm{pol} < 0.9$, and a remaining $1.2 \, \mathrm{MW}$ is radiated in the confined edge region, mostly by the XPR. 

The simulation domain is bounded in the radial direction by flux surfaces $\rho_\mathrm{pol} \in [0.9, 1.04]$.
The numerical grid spans 16 poloidal planes connected in parallel direction per the FCI approach, and each plane comprises a Cartesian grid with a resolution of $1.44 \, \mathrm{mm}$ per grid point.
Whenever simulation data is shown in the following sections, unless explicitly stated to be instantaneous, it has been averaged over toroidal angle $\varphi$ (over all 16 planes) and time $t$ (over the last 50 snapshots, spanning $0.1 \, \mathrm{ms}$). 
\blue{Instantaneous snapshots are taken at plane 4, corresponding to $\varphi = \pi / 2$.}

We apply asymptotic free-streaming limits to the parallel Braginskii heat flux of electrons and ions, $\mathrm{f}^\mathrm{FS}_\mathrm{e,i} n \sqrt{T_\mathrm{e,i} / m_\mathrm{e,i}} T_\mathrm{e,i}$ \cite{zholobenko2024}, selecting limiter coefficients $\mathrm{f}^\mathrm{FS}_\mathrm{e} = 0.3$ and $\mathrm{f}^\mathrm{FS}_\mathrm{i}  = 1$ respectively. 
Boundary conditions in perpendicular directions are set to zero-flux, i.e.~Neumann boundary conditions. At the divertor targets, we apply insulating sheath boundary conditions identically to \cite{eder2025}.
Adaptively flux-driven sources near the core boundary at $\rho_\mathrm{pol} \in [0.90, 0.912]$ damp density and temperatures to target values prescribed by corresponding \blue{experimental data (obtained from Integrated Data Analysis \cite{fischer2010})}, $n^\mathrm{core} = 3.8 \times 10^{19}\,\mathrm{m^{-3}}, T_\mathrm{e}^\mathrm{core} = T_\mathrm{i}^\mathrm{core} = 300 \, \mathrm{eV}$. 
These core sources $S^\mathrm{core}_\mathrm{n}, S^\mathrm{core}_{T_\mathrm{e}}, S^\mathrm{core}_{T_\mathrm{i}}$ thus determine the effective heating input, $P_\mathrm{heating} = \int 1.5 \, n \left( S^\mathrm{core}_{T_\mathrm{e}} + S^\mathrm{core}_{T_\mathrm{i}} \right) + 1.5 \, S^\mathrm{core}_n \left( T_\mathrm{e} + T_\mathrm{i} \right) \mathrm{d}V$.
At the target plates, we set neutrals to zero parallel velocity, $v_\mathrm{N \parallel} = 0$, and wall temperature, $T_\mathrm{N} \sim 0$. Neutral gas density at the target is set to a fixed value $N_\mathrm{div} = 1\times 10^{19} \, \mathrm{m^{-3}}$.
Note that neutrals at the boundary can also be set self-consistently using recycling boundary conditions \cite{eder2025}. 
However, we ultimately found it beneficial to apply fixed boundaries to retain more control over neutrals density in the divertor, as they play a key role in facilitating the XPR state (as will be discussed below in Section \ref{section:application:initiation}). 
Impurity concentration is set to $c_\mathrm{imp} = 10\%$ after scanning over multiple simulations from $c_\mathrm{imp} = 2\% - 10\%$
\red{and the temperature threshold parameter $T_\mathrm{N, min}$ (see equation \eqref{eq:diffusion_limiter}) is set to $30\,\mathrm{eV}$. In one particular simulation, these two parameters have been modified, as will be discussed in section \ref{section:application:height}.}


\subsection{Achieving X-point radiation}
\label{section:application:initiation}

It is important at this point to properly define the objective of this study. In AUG discharges, the XPR typically develops on the timescale of $\approx 20\,\mathrm{ms}$ after initiating the impurity puff. GRILLIX, as a turbulence code, aims to model quasi-steady states on the timescale of turbulent transport ($\approx 1\,\mathrm{ms}$). The plasma state is nonetheless susceptible to global transport time scales set by the confinement time ($\approx 10\,\mathrm{ms}$), representing an upper time limit on the quasi-steady turbulent state. Therefore, attempting to capture the full onset of the XPR regime is beyond the scope of this work, and we instead focus on the study of plasma turbulence during an established XPR state. 

\begin{figure}
    \centering
    \includegraphics[width=0.99\linewidth,trim={0 32 0 0},clip]{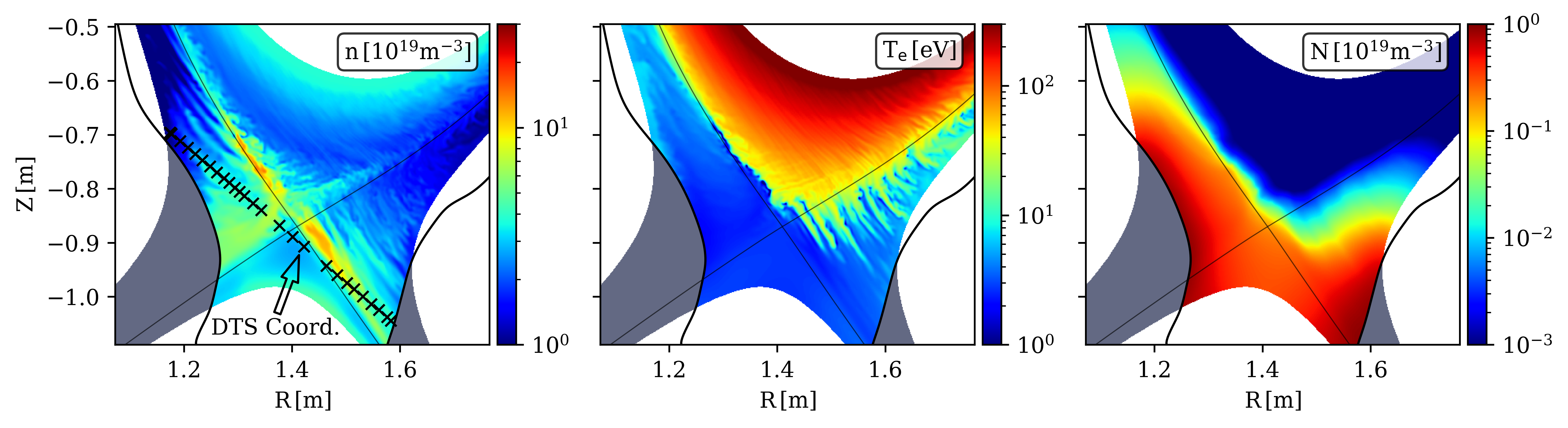}
    \includegraphics[width=0.99\linewidth,trim={0 32 0 0},clip]{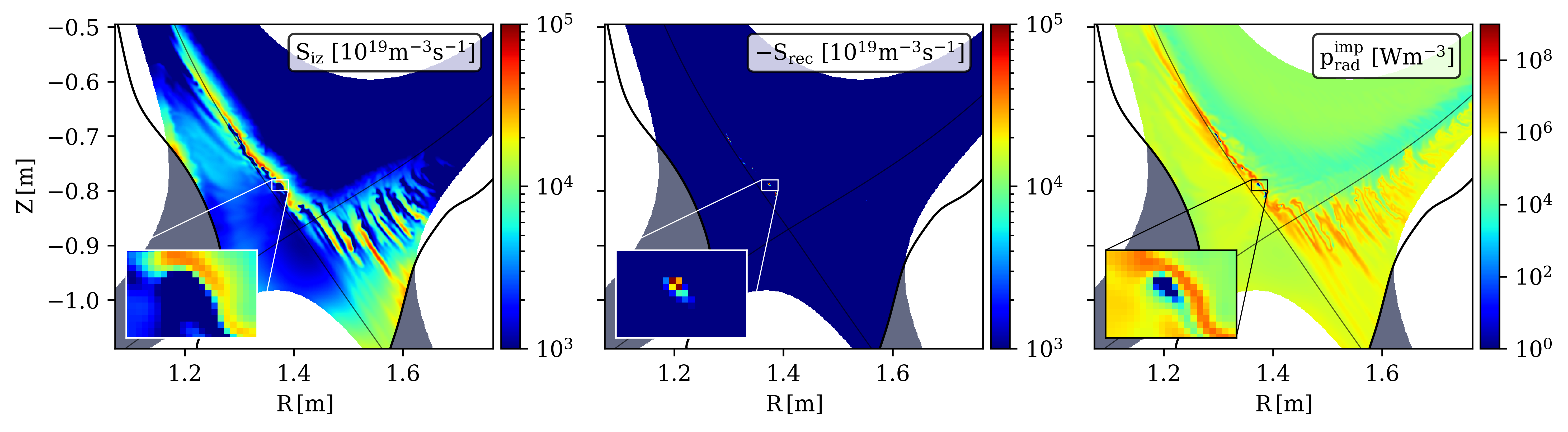}
    \includegraphics[width=0.99\linewidth,trim={0 5 0 0},clip]{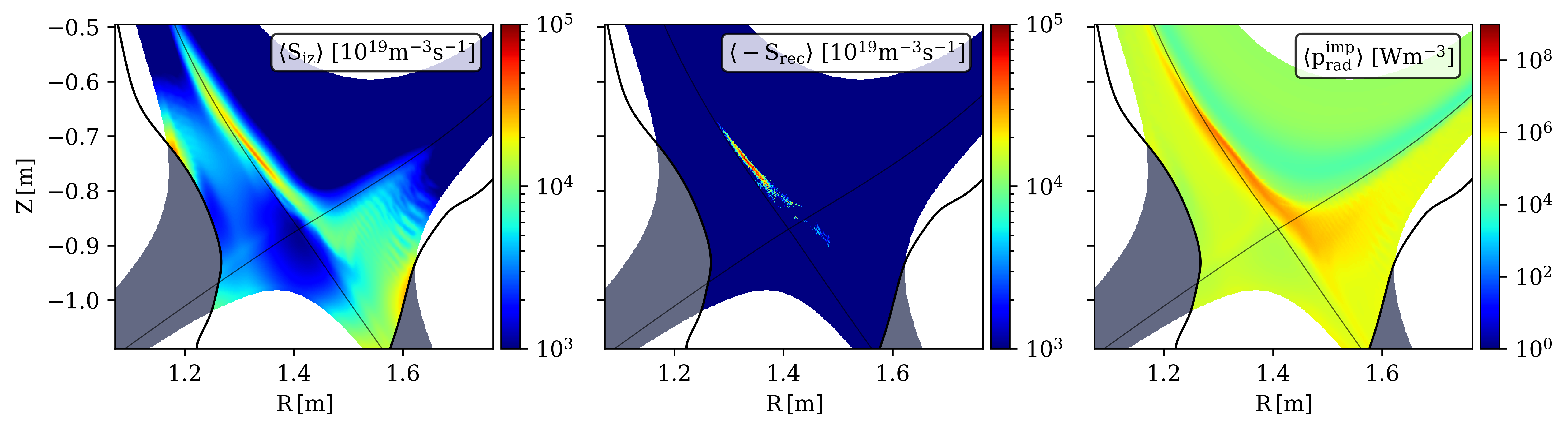}
    \caption{Snapshots of various plasma quantities in a simulation with X-point radiator present (later referred to as ``Low-XPR'' case) \blue{taken at the $\varphi = \pi / 2$ plane}. First row: instantaneous plasma density, electron temperature, and neutrals density. The line-of-sight of the Divertor Thomson Scattering (DTS) diagnostic shown in Figure \ref{fig:dts_profiles} is marked with purple crosses. Second row: instantaneous ionization rate, recombination rate, and impurity radiation density. Third row: same as second row but averaged over time $t$ and toroidal angle $\varphi$.}
    \label{fig:lo_xpr_snap}
\end{figure}

X-point radiation in simulations was achieved by introducing, at initial stages of the simulation (when initial profiles develop into turbulence) an additional, temporary source of neutral particles with $T_\mathrm{N} \approx 0\,\mathrm{eV}$ in the divertor \footnote{Sourcing particles over time is preferable to directly setting higher initial/boundary neutrals density, as the latter might induce strong local ionization and numerically unstable gradients.}.
The additional source $S_\mathrm{N}$ is defined as torodially symmetric 2D-Gaussian, centered at $R = 1.38\,\mathrm{m}, Z = -0.85\,\mathrm{m}$, with a Gaussian width $\sigma = 10\,\mathrm{cm}$, and generating a total of $1.8\times10^{19}\,\mathrm{D}$ over $0.5\,\mathrm{ms}$.
This helps cool down the divertor and X-point until local impurity radiation becomes sufficient to keep temperatures low and a radiating front develops. 
Crucially, after the removal of the temporary source, the radiation front persists.
Figure \ref{fig:lo_xpr_snap} depicts the simulation state following the establishment of a stable radiation front. At this point, the additional neutrals source has been disabled for $\approx 1\,\mathrm{ms}$.

We identify a cold, dense structure above the X-point toward the high-field side, with plasma density (first row, left) exceeding the surrounding area by an order of magnitude ($\approx 20\times10^{19}\,\mathrm{m^{-3}}$ vs. $3\times10^{19}\,\mathrm{m^{-3}}$). Below the dense XPR structure, electron temperature $T_\mathrm{e}$ (first row, center) drops to single-digit electron volts, enabling neutral gas (first row, right) to reach the confined region before being ionized. At the low-field side, by contrast, plasma near the separatrix is significantly hotter, and neutrals density is correspondingly lower.
The intersection of confined hot plasma and the neutrals cloud is marked by a front of ionization and radiation (second row). The turbulent radiative front surrounds small pockets of dense, sub-$1\,\mathrm{eV}$ plasma, in which recombination becomes significant. Although these spots may appear tiny, they provide a significant plasma particle sink in the XPR core when averaged toroidally and in time (third row). 
While the intermittency of ionization and recombination rates in turbulence simulations has been pointed out previously \cite{fan2019, eder2025}, we reiterate that the rate coefficients depend on plasma density and electron temperature non-linearly. As a consequence, the averaged particle transfer rates in highly fluctuating turbulence likely differ from those obtained in mean field simulation. This will be investigated more thoroughly in future work.
A trivial but important finding from our initiation attempts is that impurity concentration must be sufficiently high for the XPR to persist. Reducing $c_\mathrm{imp}$ from $10\%$ to $2\%$, for example, while not destroying the XPR outright, leads to the ionization/radiation structure gradually drifting below the separatrix, after which it vanishes entirely. 

\subsection{Height modulation}
\label{section:application:height}

\begin{figure}
    \centering
    \includegraphics[width=0.9\linewidth,trim={0 10 0 5},clip]{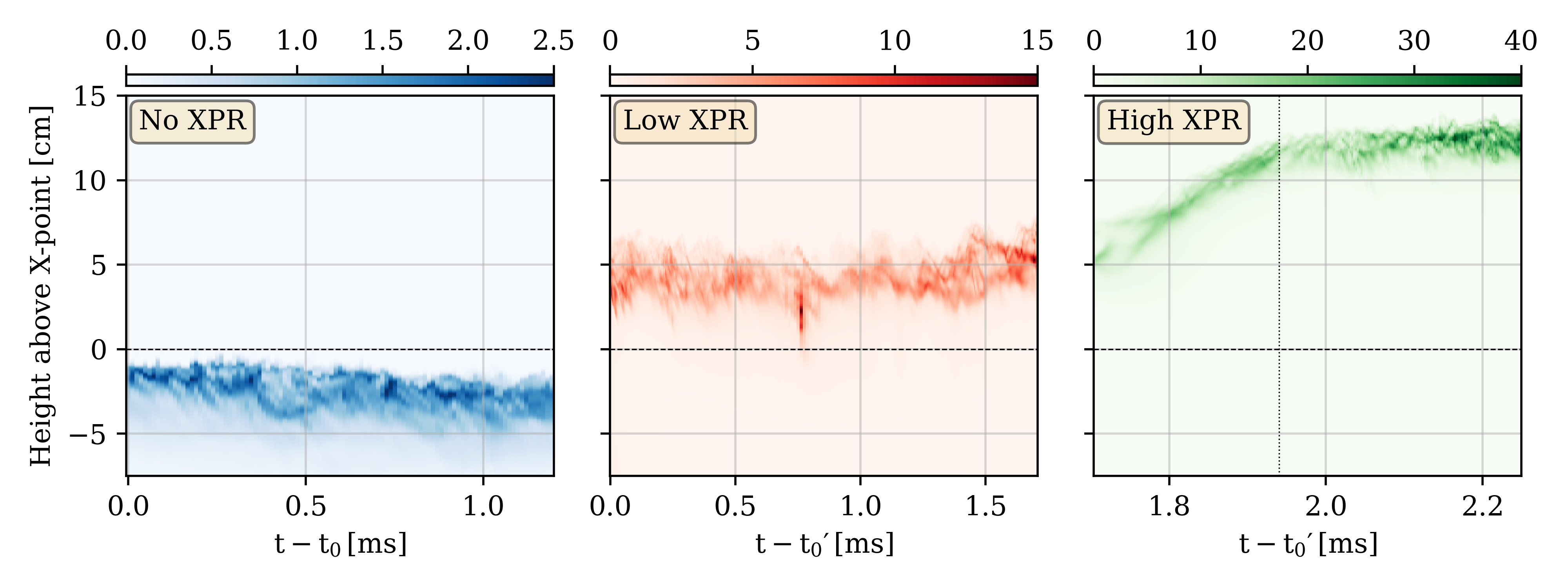}
    \caption{Impurity radiation density $p_\mathrm{rad}^\mathrm{imp} \, [\mathrm{MW m^{-3}}]$ directly above the X-point as a function of simulation time. Note that the subplots apply different upper limits of 2.5, 15, and 40 $\mathrm{MW m^{-3}}$. The High-XPR simulation begins at the final state of the Low-XPR simulation at $1.7\,\mathrm{ms}$, where the divertor neutrals density $N_\mathrm{div}$ is increased. At $1.94\,\mathrm{ms}$, divertor neutrals density is reduced back to $1\times10^{19}\,\mathrm{m^{-3}}$ and the radiation height stabilizes.}
    \label{fig:xpr_height}
\end{figure}

In AUG experiments, \blue{the XPR is known to be controllable up to a height of $15\,\mathrm{cm}$ above the X-point \cite{bernert2021}, and is estimated to be $\approx 15\,\mathrm{cm}$ above the X-point at the time of interest $t = 2.4\,\mathrm{s}$ during discharge \#40333 \cite{bernert2025}}. 
In simulations, we estimate the height of the XPR structure by plotting the impurity radiation density vertically above the X-point, shown in Figure \ref{fig:xpr_height}. The XPR simulation discussed so far is in the center (denoted as ``Low XPR'' case), where we find that the radiating structure stays roughly at the same height of $ \approx 5\,\mathrm{cm}$ above the X-point. 
For reference, we also consider a simulation without X-point radiation (left subplot), referred to as ``No XPR'' case. It was performed under identical conditions, though without the initial sourcing of neutrals in the divertor as outlined in \ref{section:application:initiation}. It features a radiative front as well, although it is located below the X-point and is significantly weaker.

\begin{figure}
    \centering
    \includegraphics[width=0.99\linewidth,trim={0 32 0 5},clip]{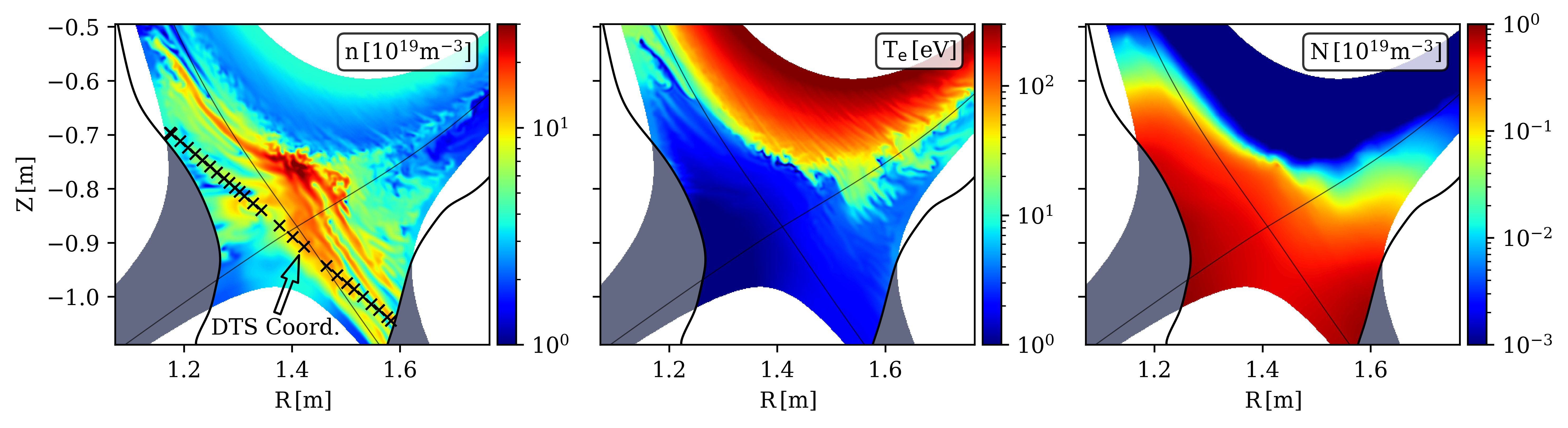}
    \includegraphics[width=0.99\linewidth,trim={0 32 0 5},clip]{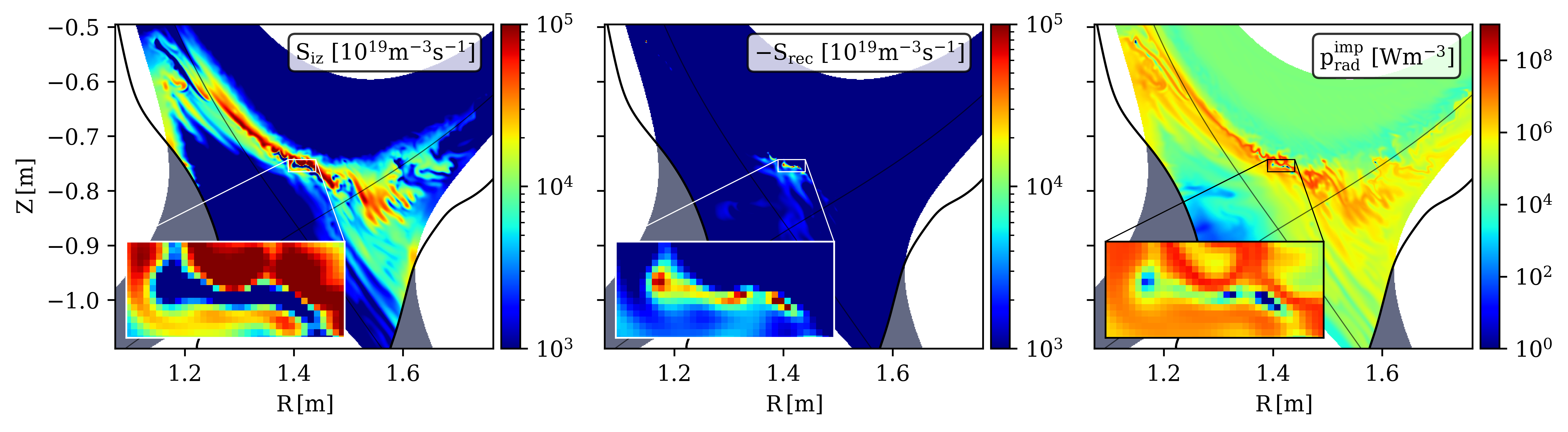}
    \includegraphics[width=0.99\linewidth,trim={0 5 0 5},clip]{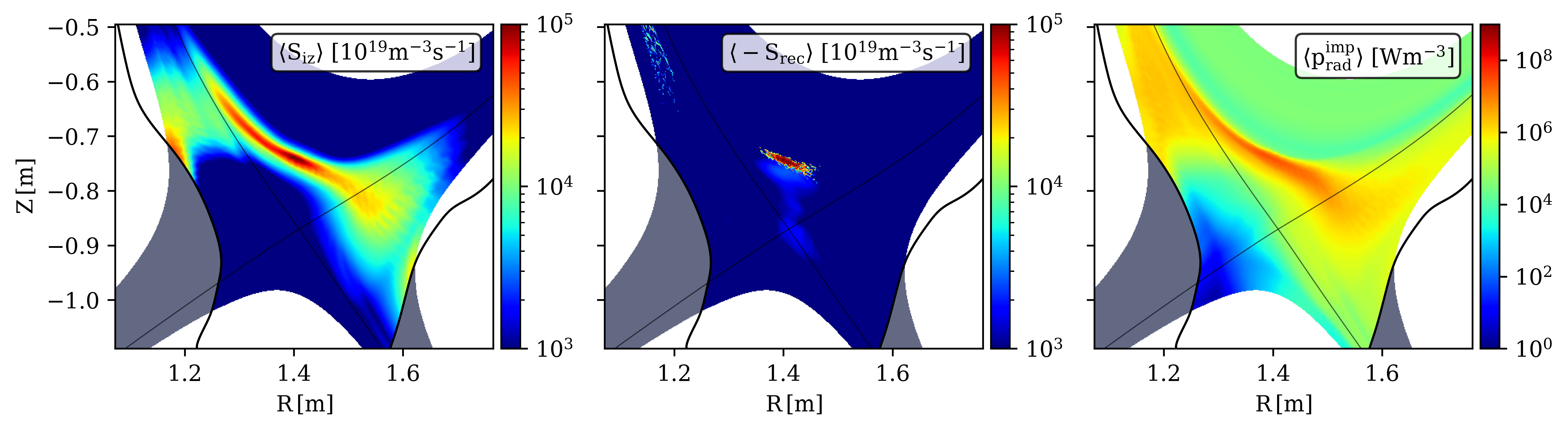}
    \caption{\blue{Same as Figure \ref{fig:lo_xpr_snap} for the High-XPR simulation.}}
    \label{fig:hi_xpr_snap}
\end{figure}

The right subplot in Figure \ref{fig:xpr_height} shows a third simulation, in which the radiation front could be pushed to $\approx 12\,\mathrm{cm}$ (denoted ``High XPR'' case). This was achieved by continuing the ``Low XPR" simulation from $1.7\,\mathrm{ms}$ with higher divertor neutrals density $N_\mathrm{div} = 1\times10^{19} \rightarrow 2\times10^{19}\,\mathrm{m^{-3}}$ and enforcing a minimum diffusivity $T_\mathrm{N, min} = \red{30\rightarrow \,} 50\,\mathrm{eV}$.
This results in flooding the divertor with additional neutral gas, pushing the XPR structure upwards, see Figure \ref{fig:hi_xpr_snap}. After $0.24\,\mathrm{ms}$, the divertor neutrals density is returned to its original value $N_\mathrm{div} = 1\times10^{19}\,\mathrm{m^{-3}}$. Concurrently, the impurity concentration parameter was reduced from $c_\mathrm{imp} = 10\% \rightarrow 5\%$. This was found to prevent the radiation front from rising too high (towards $\rho_\mathrm{pol} \approx 0.98$).

In this "High XPR" simulation, we note that the formerly identified characteristics of the X-point radiating state become more pronounced (see Figure \ref{fig:hi_xpr_snap}). Plasma density in the XPR core rises by another $\approx 50\%$, the divertor cools to below $2\,\mathrm{eV}$, and neutrals density at the X-point increases to $\approx 1\times10^{19}\,\mathrm{m^{-3}}$.
Correspondingly, the ionization and radiation fronts rise to $12\,\mathrm{cm}$ above the X-point, enveloping intermittent cold cores where plasma recombines.

\begin{figure}
    \centering
    \includegraphics[width=0.45\linewidth,trim={0 5 0 5},clip]{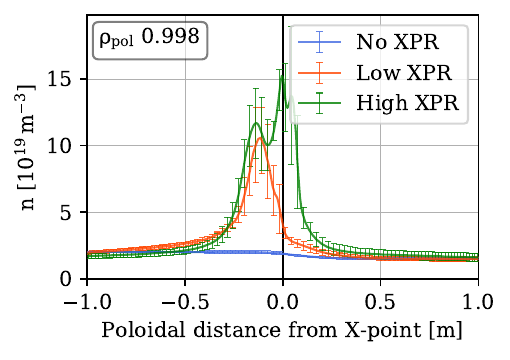}
    \includegraphics[width=0.45\linewidth,trim={0 5 0 5},clip]{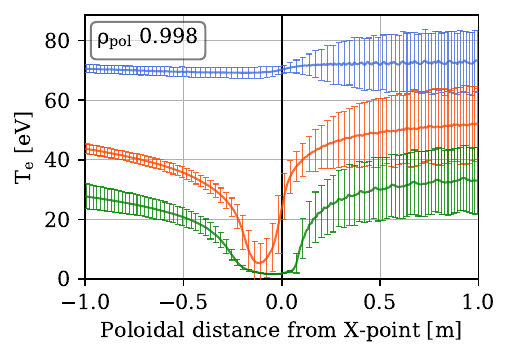}
    \caption{Plasma density and electron temperature on flux surface $\rho_\mathrm{pol}=0.998$ near the X-point. The error bars indicate fluctuation amplitudes. The flux surface passes through the recombining center of the Low-XPR simulation and passes below the XPR in the High-XPR simulation.}
    \label{fig:poloidal_profiles}
\end{figure}

Figure \ref{fig:poloidal_profiles} shows density and temperature on flux surface $\rho_\mathrm{pol} = 0.998$ in all simulations. The solid central lines show the mean over time and toroidal angle, and error bars indicate one standard deviation from the mean (thus providing a measure of fluctuation amplitude).
It approximately intersects with the averaged XPR core in the Low-XPR simulation, as seen in the local density peak and temperature trough close to the X-point. Although the average temperature remains above $10\,\mathrm{eV}$, large error bars indicate elevated fluctuation amplitudes, leading to the intermittent occurrence of cold recombination cores. In the No-XPR reference, it is evident that no significant profile change occurs near the X-point. 
For the High-XPR simulation, the chosen flux surface $\rho = 0.998$ passes below the dense XPR core (which resides at $\rho_\mathrm{pol} \sim 0.989$), where electron temperature remains consistently below $3\,\mathrm{eV}$ without significant fluctuations. 
Regardless of XPR conditions, we can identify ballooned transport on the outboard side \cite{zholobenko2023} by heightened fluctuation amplitudes in temperature. This also holds for density, although it is overshadowed by the dramatic amplification of density with XPR present.  

\section{Validation}
\label{section:validation}

Having verified the presence of X-point radiation in the simulations, let us now quantify their agreement with experimental measurements. A comparison of approximate heating and radiated powers is shown in Table \ref{tab:heating_power}. The total experimental heating input of $1.7\,\mathrm{MW}$ is adjusted to exclude $0.4\,\mathrm{MW}$ of radiation in the core region $\rho_\mathrm{pol} < 0.9$, which is not captured in the simulation domain. 
Simulated heating power is obtained by measuring the activity of the adaptively flux-driven core sources as outlined in Section \ref{section:application:setup}.
We thus find that the heating powers in the two XPR simulations are higher than in the experiment by 20\% and 50\%, respectively. However, note that both simulations feature correspondingly high impurity radiation, bringing the total radiation fraction close to the experimental value of 0.9.

\begin{table}
\centering
\begin{tabular}{l|l|l|l}
                                       & $P_\mathrm{heating}$ $[\mathrm{MW}]$  & $P_\mathrm{rad}$ $[\mathrm{MW}]$    & $P_\mathrm{rad} / P_\mathrm{heating}$ \\ \hline
Experiment ($\rho_\mathrm{pol} > 0.9$) & 1.3                        & 1.2                           & 0.9   \\
"No XPR" simulation                    & 1.2                        & 0.2                           & 0.2  \\ 
"Low XPR" simulation                   & 1.6                        & 1.3                           & 0.8   \\
"High XPR" simulation                  & 2.0                        & 1.6                           & 0.8   \\
\end{tabular}
\caption{Approximate heating power, radiated power, and radiation fraction in the experiment and simulations. We exclude the core $\rho_\mathrm{pol} < 0.9$ from the experimental measurement, as it is not represented in the simulation domain.}
\label{tab:heating_power}
\end{table}

\subsection{Outboard-midplane profiles}

\begin{figure}
    \centering
    \includegraphics[width=0.4\linewidth,trim={0 12 0 10},clip]{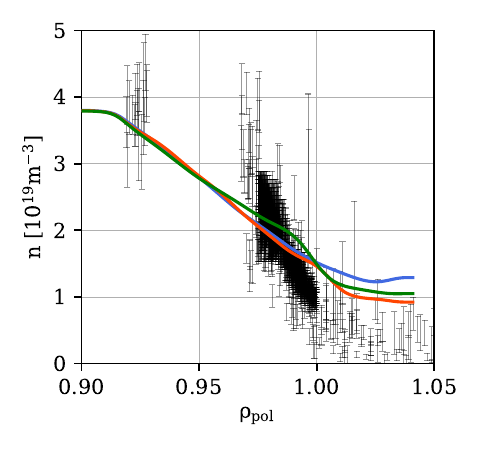}
    \includegraphics[width=0.4\linewidth,trim={0 12 0 10},clip]{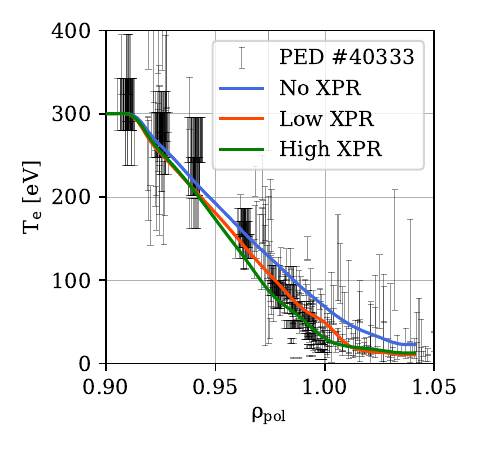}
    \caption{OMP profiles of plasma density (left) and electron temperature (right) compared to experiment data \#40333 PED $2.3-2.5\,\mathrm{s}$.}
    \label{fig:omp_profiles}
\end{figure}

A comparison of OMP profiles with experiment data (\#40333 PED, $2.3-2.5\,\mathrm{s}$) is shown in Figure \ref{fig:omp_profiles}. The left subplot shows plasma density $n$ with experimental data obtained from lithium beam and Thomson scattering diagnostics. 
Experimental data is not available continuously along the region of interest, thus a direct comparison is challenging. However, it is evident that density in the SOL $\rho_\mathrm{pol} \in [1.0,1.04]$) is overestimated across simulations, leveling out at $1\times10^{19}\,\mathrm{m^{-3}}$ compared to $0.4\times10^{19}\,\mathrm{m^{-3}}$ in the experiment.
The right subplot compares electron temperature $T_\mathrm{e}$ against electron cyclotron emission and Thomson scattering measurements. The profile is matched very well, including in the outer SOL. The High-XPR case yields the best agreement. 
 
\subsection{Divertor profiles}

\begin{figure}
    \centering
    \includegraphics[width=0.8\linewidth,trim={0 4 0 5},clip]{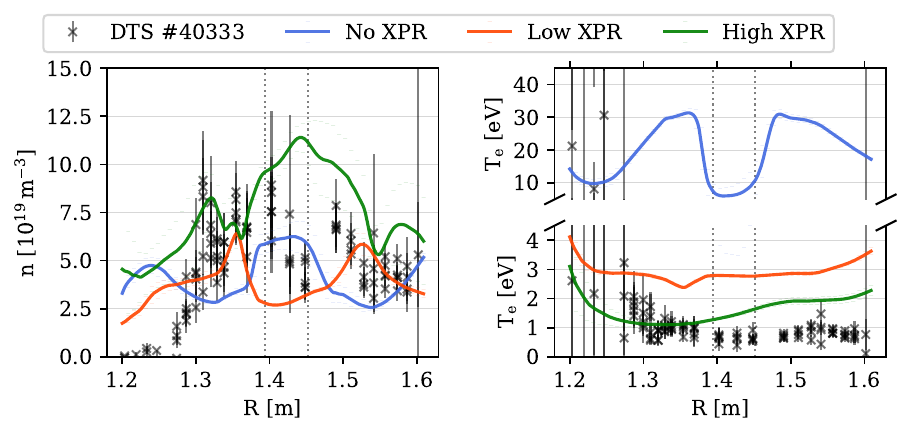}
    \caption{Experimental measurements \#40333 DTS $2.3-2.5\,\mathrm{s}$ overlaid with simulation data for plasma density (left) and electron temperature (right). \blue{Dotted vertical lines indicate where the separatrix intersects with the line of DTS measurement coordinates (shown in the top left subplots of Figures \ref{fig:lo_xpr_snap} and \ref{fig:hi_xpr_snap}).}}
    \label{fig:dts_profiles}
\end{figure}

A similar comparison is performed below the X-point, where density and electron temperature are plotted against Divertor Thomson Scattering measurements (\#40333 DTS, $2.3-2.5\,\mathrm{s}$); see Figure \ref{fig:dts_profiles}. The DTS measurement locations are marked in the left subplots of Figures \ref{fig:lo_xpr_snap}, \ref{fig:hi_xpr_snap}. They are arranged in a line, onto which the simulation data is interpolated.
Comparing plasma density (left subplot), we once more find approximate agreement. While all simulations roughly cluster around the experimental values, the No-XPR and Low-XPR cases tend to undershoot, whereas the High-XPR case overestimates the measurements.
The comparison of electron temperature (right subplot) is significantly easier to judge. The No-XPR reference case is still attached, featuring temperatures of up to $30\,\mathrm{eV}$ compared to $\approx 1\,\mathrm{eV}$ in the experiment. The Low-XPR case performs markedly better, yielding values in the same order of magnitude at $3 - 4\,\mathrm{eV}$. Finally, the High-XPR case provides the closest match at temperatures of $1 - 2\,\mathrm{eV}$.
\blue{Note that plasma in the experiment is mostly absent at $R < 1.27\,\mathrm{m}$, resulting in large uncertainties of electron temperature measurements due to low signal strength.}

\subsection{Bolometry}

\begin{figure}
    \centering
    \includegraphics[width=0.6\linewidth,trim={0 5 0 5},clip]{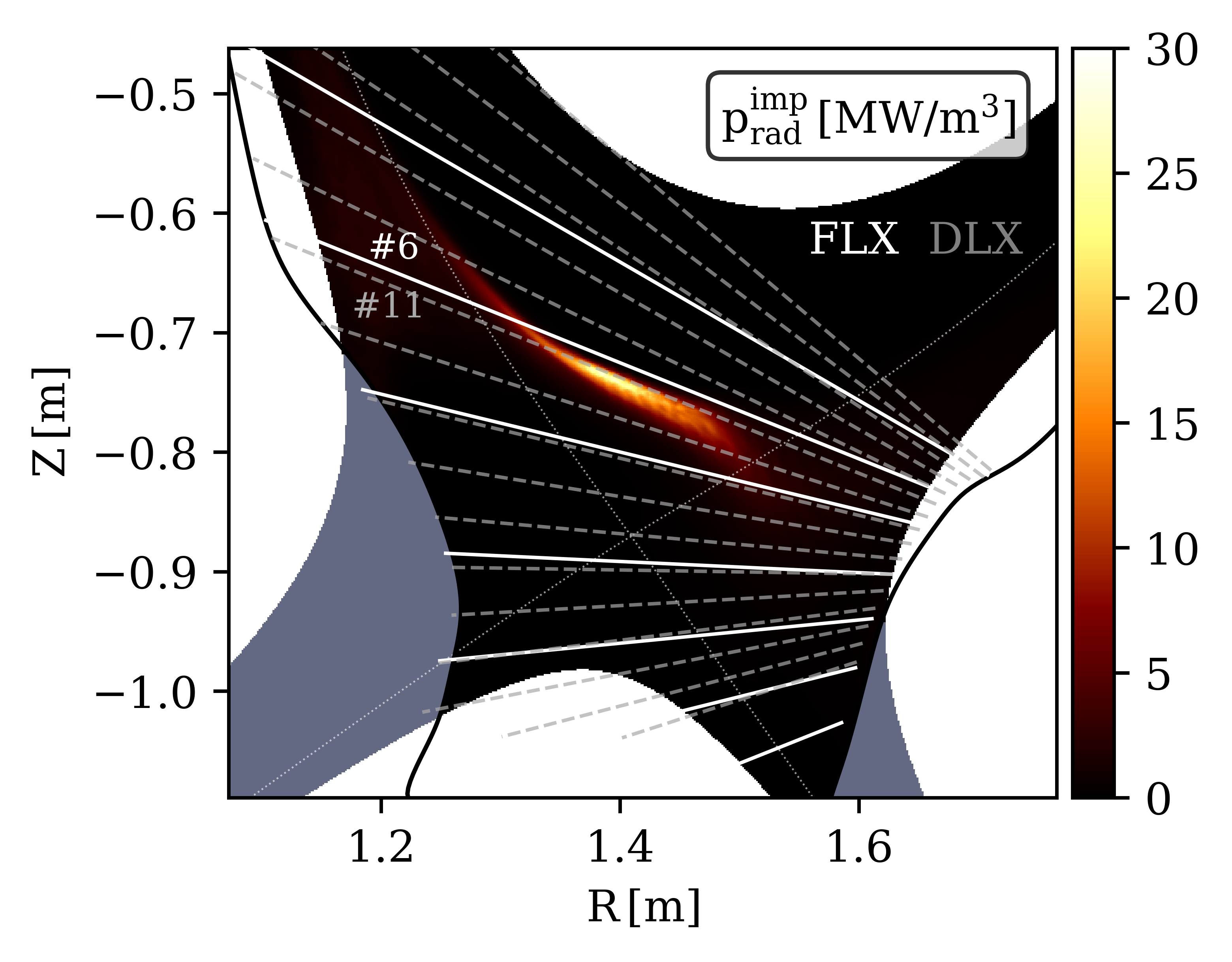}
    \caption{Bolometry sight lines in the experiment overlaid on impurity radiation density in the "High XPR" simulation, FLX \#1-7 in white solid lines, DLX \#1-16 in gray dashed lines (increasing from bottom to top). The channels closest to the simulated radiation peak (FLX \#6 and DLX \#11) are numbered explicitly.}
    \label{fig:bolometry_channels}
\end{figure}

\begin{figure}
    \centering
    \includegraphics[width=0.49\linewidth,trim={0 5 0 5},clip]{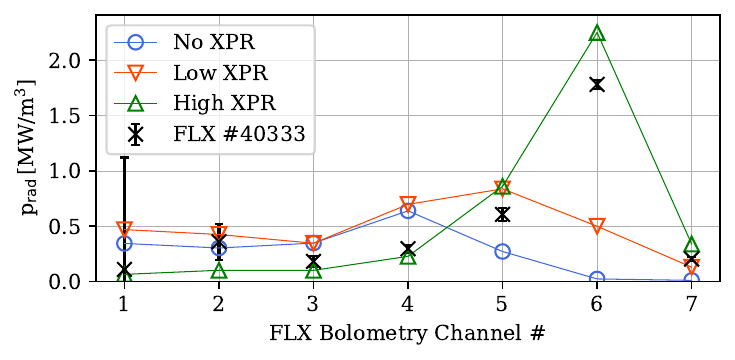}
    \includegraphics[width=0.49\linewidth,trim={0 5 0 5},clip]{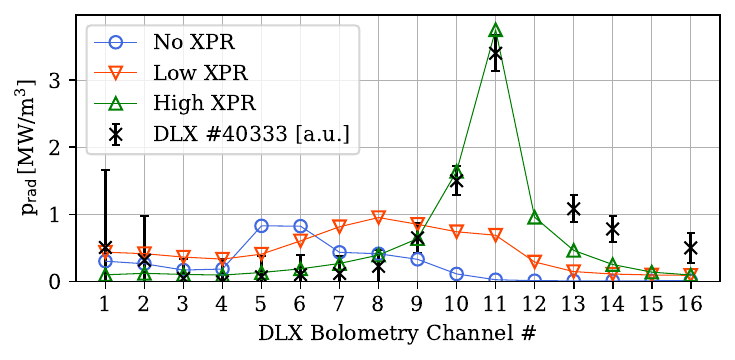}
    \caption{Comparison between radiation in simulations and experimental measurements of  \#40333 FLX $2.38 - 2.42\,\mathrm{s}$ (left) and \#40333 DLX $2.399 - 2.401\,\mathrm{s}$ (right). Simulation data and FLX measurements are shown in physical units. DLX signal strength is in arbitrary units and scaled to approximately match the peak value of the High-XPR simulation. No data is available for DLX \#12 and \#15.}
    \label{fig:bolometry_comparison}
\end{figure}

Finally, let us quantify how the radiating structures found in the simulations compare to the experiment. For this purpose, we consider the horizontal lower divertor viewing cameras from foil bolometry (\#40333 FLX, $2.38-2.42\,\mathrm{s}$) and diode bolometry (\#40333 DLX, $2.399-2.401\,\mathrm{s}$). Figure \ref{fig:bolometry_channels} shows the respective sight lines of both cameras, 7 for FLX, and 16 for DLX (with channels \#12 and \#15 not functioning).
We average simulated radiation density toroidally and over the area associated with each sight line, defined as spanning halfway to the sight lines below and above it.
Figure \ref{fig:bolometry_comparison} shows the resulting power density per channel (FLX on the left, DLX on the right).
The DLX camera features higher spatial resolution compared to FLX, and is used for real-time control of the XPR height during the experiment \cite{bernert2021}.
However, its signal is not convertible to physical units \cite{bernert2014axuv}. We therefore scale it such that the experimental peak aligns approximately with the peak value obtained in the High-XPR case. 
It is nonetheless obvious that the High-XPR case agrees well with both FLX and DLX bolometry profiles. 
Both experiment and simulation feature negligible radiation at low channels, which then sharply ramps to a peak around FLX \#6 and DLX \#11.
By comparison, radiation in the No-XPR and Low-XPR simulations is distributed much more broadly and occurs at lower channels.
\red{The agreement of the High-XPR case is remarkable, also in absolute values (though note that some overestimation should be expected due to higher radiation compared to the experiment, see Table \ref{tab:heating_power}).}

\section{Turbulence and transport analysis}
\label{section:analysis}

\subsection{Poloidal structure of the X-point radiator}

\begin{figure}
    \centering
    \includegraphics[width=0.7\linewidth,trim={0 5 0 5},clip]{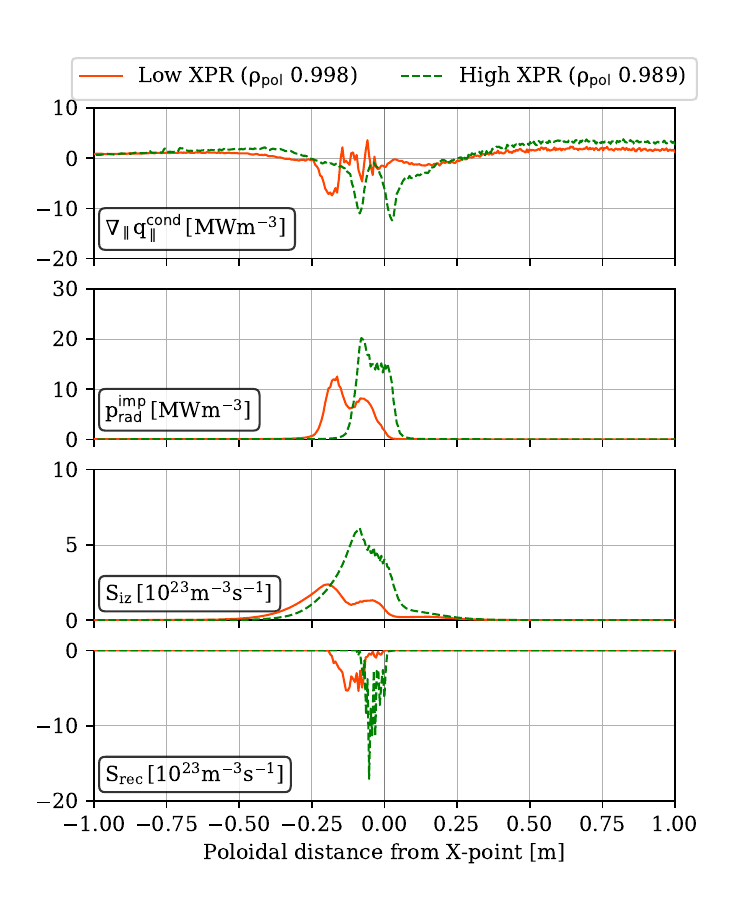}
    \caption{Poloidal projection of power sink by heat conduction (1st row), impurity radiation density (2nd row), ionization rate (3rd row), and recombination rate (4th row). The chosen flux surfaces $\rho_\mathrm{pol} = 0.998, 0.989$ intersect with the recombination-dominated core in the Low-XPR and High-XPR simulations.}
    \label{fig:xpr_poloidal}
\end{figure}

As briefly mentioned in \ref{section:model:impurity}, earlier analytical and transport studies \cite{stroth2022xpr, pan2022} proposed that the XPR represents a stable power balance between heat conducted into the XPR volume and radiation leaving the volume.  
To investigate this hypothesis, Figure \ref{fig:xpr_poloidal} shows a poloidal projection of various quantities in the XPR simulations: 
the parallel conductive heat source/sink $\nabla \cdot q_\parallel^\mathrm{cond} \mathbf{b}$, radiation density $p_\mathrm{rad}^\mathrm{imp}$, ionization rate $S_\mathrm{iz}$, and recombination rate $S_\mathrm{rec}$. The flux surfaces to project onto are chosen such that they intersect the cold XPR in each simulation and are additionally marked in Figure \ref{fig:2d_fluctuation} (as will be discussed further in the following paragraphs).
Figure \ref{fig:xpr_poloidal} reveals a conductive heat sink ($\nabla \cdot q_\parallel^\mathrm{cond} < 0$) forming in the XPR volume (predominantly by electrons), which aligns with the onset of impurity radiation.
It also coincides with a sharp increase in ionization rate, which surrounds a narrower recombination area in the center of the XPR volume.
Note that, due to the turbulent nature of the system, there is no clear delineation between regions of ionization and recombination when averaged over time. 
Their spatial separation becomes apparent only when observing the instantaneous plasma state (see Figures \ref{fig:lo_xpr_snap}, \ref{fig:hi_xpr_snap}), as discussed in Section \ref{section:application}.
Nonetheless, we find that the morphology of the turbulent XPR structure agrees qualitatively with previous transport studies \cite{pan2022}.

\subsection{Fluctuation amplitudes}

Next, we perform a comparative analysis across simulations to identify differences arising from the varying degrees of detachment. 
In Figure \ref{fig:histogram}, we show the probability distribution functions (PDFs) of density and electron temperature \blue{normalized to their time and toroidally averaged values, $n / \langle n \rangle$ and $T_\mathrm{e} / \langle T_\mathrm{e} \rangle$.}
To isolate the impact of XPR structures, data is again taken on the respective flux surface containing the cold recombination zone, $\rho_\mathrm{pol} = 0.998, 0.989$. Additionally, it is limited to a poloidal segment near the X-point (poloidal angle $\theta_\mathrm{pol} \in [4.08, 4.78]\,\mathrm{rad}$ with respect to the magnetic axis), marked in Figure \ref{fig:2d_fluctuation}.
\blue{The distributions of density and temperature are approximately symmetric and close to Gaussian in the No-XPR case, and widen significantly in the Low- and High-XPR cases.}
\blue{Fluctuations in the XPR cases exceed the background by up to 400 \% in density, compared to 40\% of the background in the reference.} 
\blue{Additionally, we also find a broadening of fluctuations in the opposite direction; a large population of temperature samples are found at \blue{$T_\mathrm{e} / \langle T_\mathrm{e} \rangle \approx 0.0$} in the XPR cases, indicative of the intermittent cold recombination cores.}
The broadening of fluctuations is further evident in the 2D view, as \blue{shown in Figure \ref{fig:2d_fluctuation} for plane $\varphi = \pi/2$.}
These strong fluctuations are not confined solely to the XPR region but extend into the SOL as well, highlighting the non-local nature of the underlying turbulence.
\textbf{\begin{figure}
    \centering
    \includegraphics[width=0.8\linewidth,trim={0 5 0 5},clip]{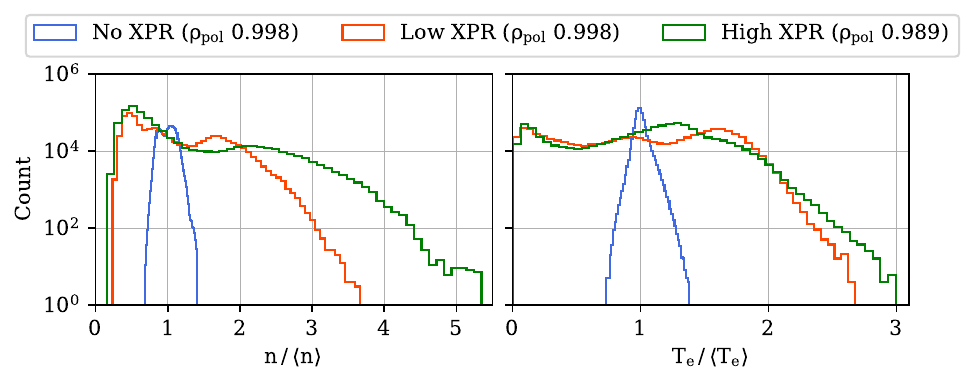}
    \caption{\blue{Histogram of density and electron temperature values (normalized to their background)} observed in the X-point sector ($\theta_\mathrm{pol} \in [4.08, 4.78]\,\mathrm{rad}$) at flux surfaces $\rho_\mathrm{pol} = [0.998, 0.998, 0.989]$ depending on simulation (see also Figure \ref{fig:2d_fluctuation}).}
    \label{fig:histogram}
\end{figure}
}
\begin{figure}
    \centering
    \includegraphics[width=0.99\linewidth,trim={0 33 0 5},clip]{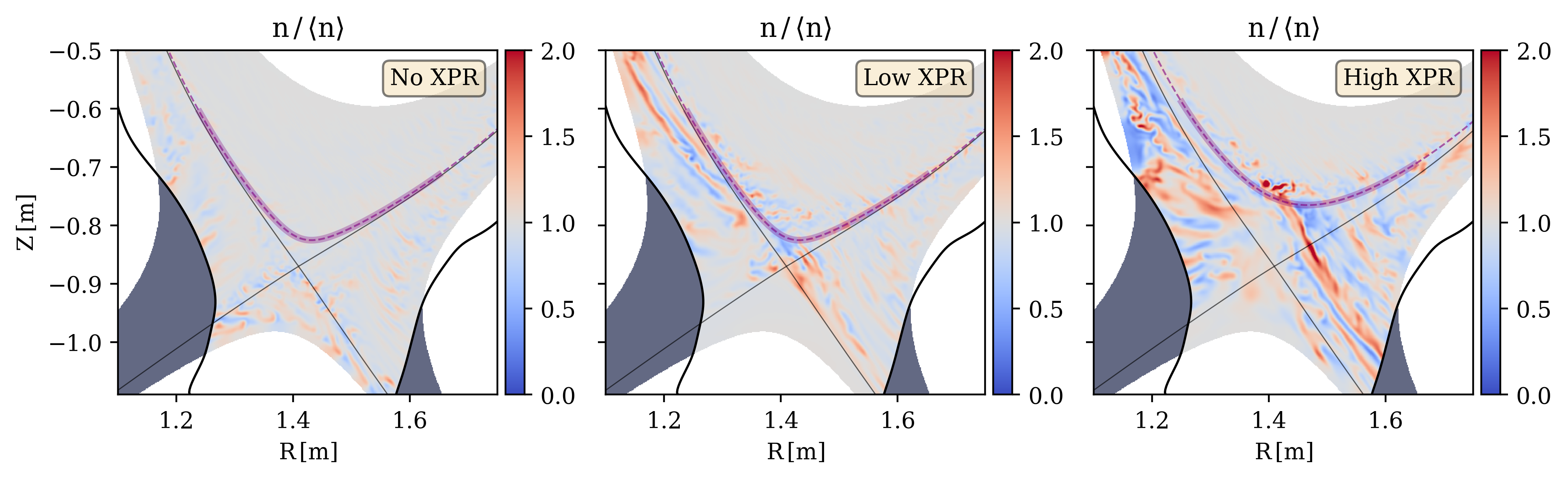}
    \includegraphics[width=0.99\linewidth,trim={0 5 0 5},clip]{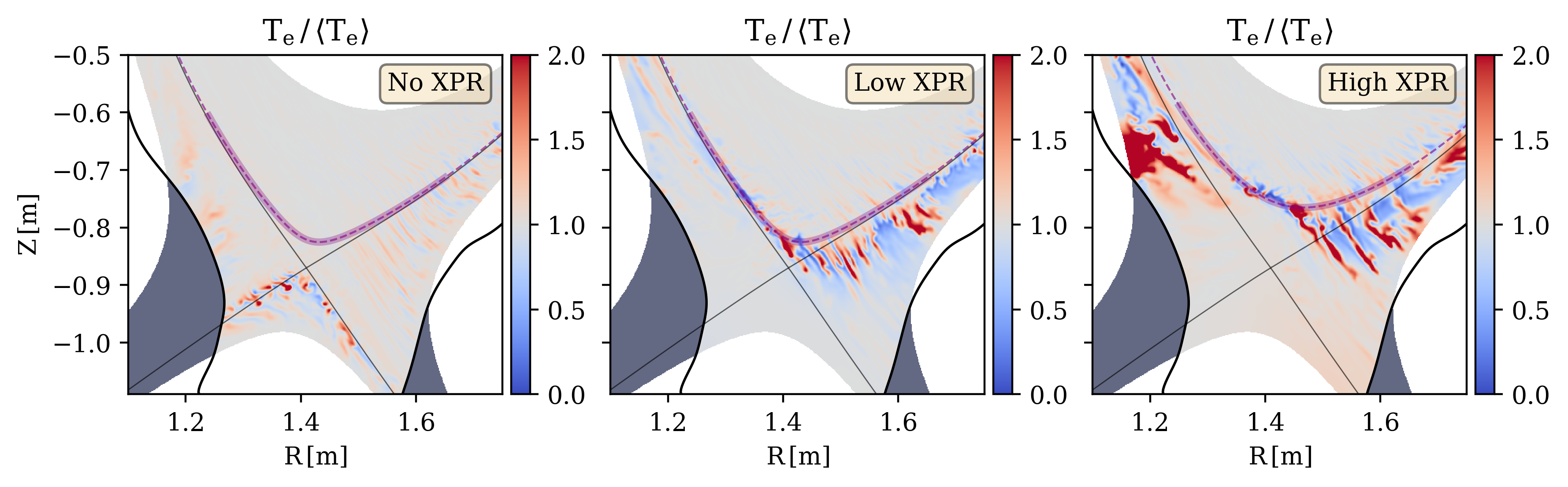}
    \caption{\blue{Instantanous snapshot of plasma density and electron temperature at plane $\varphi = \pi / 2$ normalized to their background} in the No-XPR case (left column), Low-XPR case (center column), and High-XPR case (right column). The X-point sector of the respective flux surface of interest ($\rho_\mathrm{pol} = [0.998, 0.998, 0.989]$, see Figure \ref{fig:histogram}) is highlighted in purple.}
    \label{fig:2d_fluctuation}
\end{figure}
Note also that such large fluctuations further attest to the necessity of a full-\textit{f} approach, as applied in GRILLIX.

\subsection{\red{Transport coefficients}}

\begin{figure}
    \centering
    \includegraphics[width=0.7\linewidth]{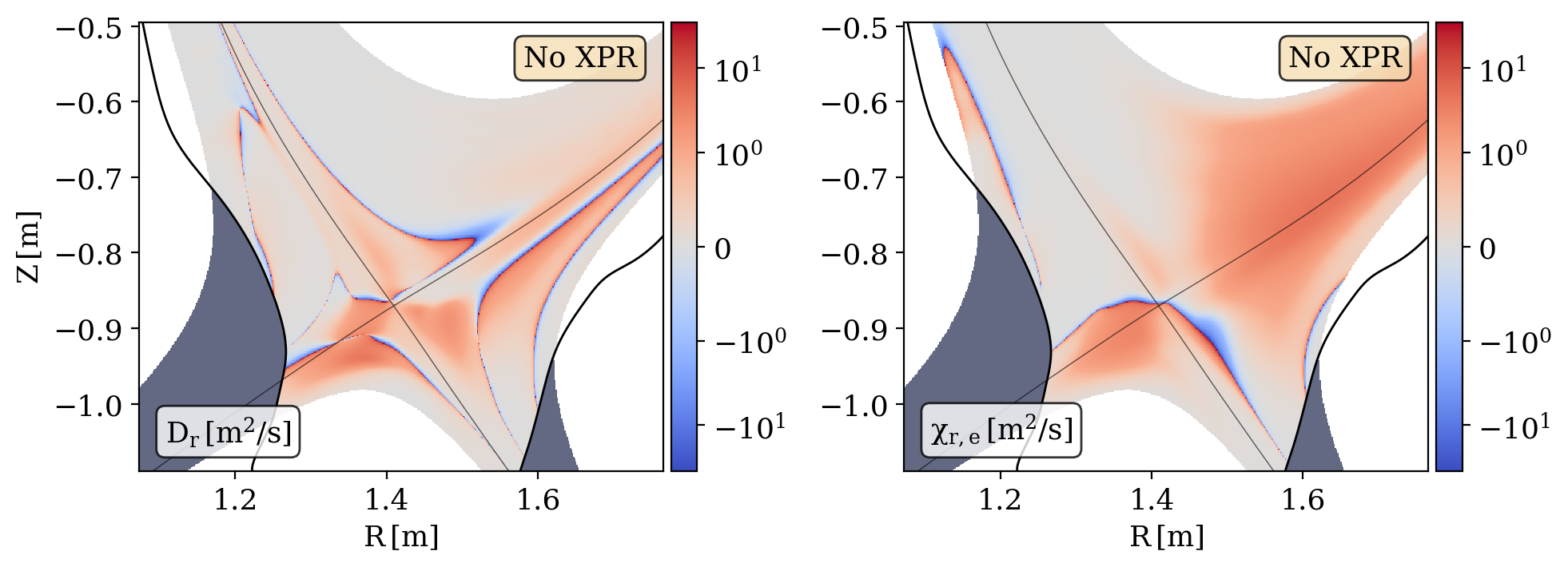}
    \includegraphics[width=0.7\linewidth]{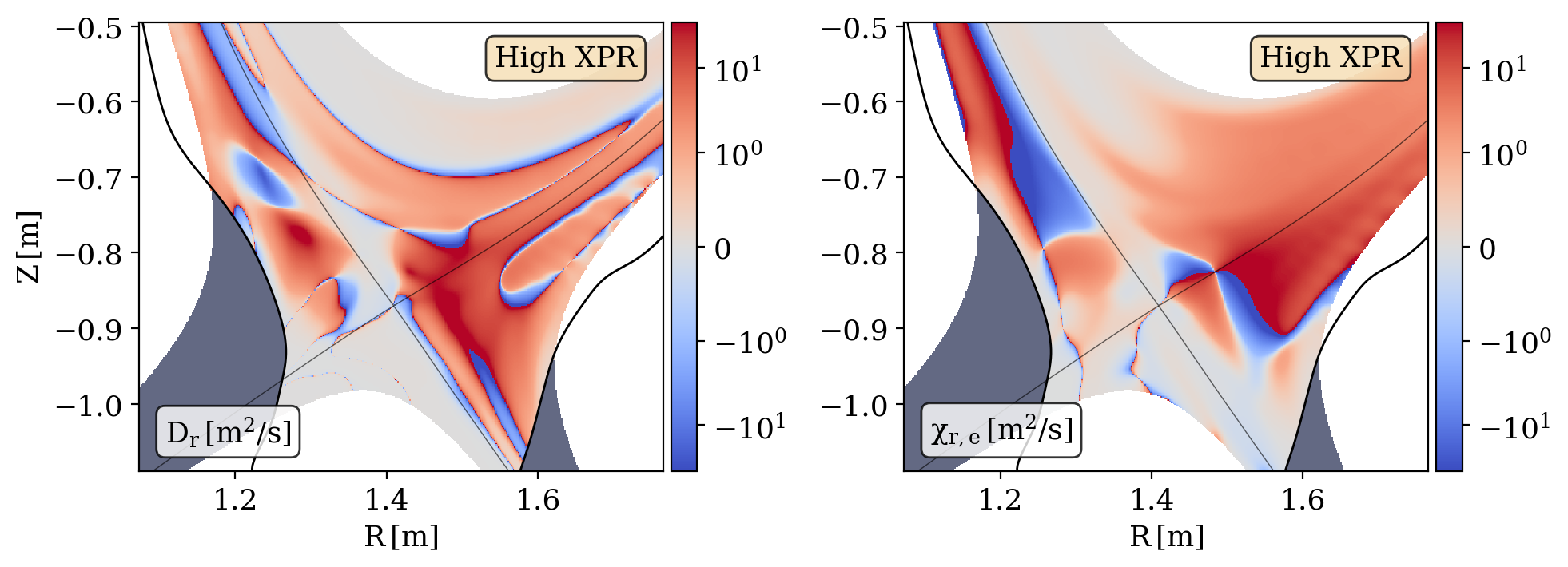}
    \caption{Radial particle diffusivity (1st column) and electron heat conductivity (2nd column) for the No-XPR case (top row) and High-XPR case. Negative values (in blue) indicate regions where radial flow points in direction of the density / temperature gradient (pinch).}
    \label{fig:diffusivities}
\end{figure}

The drastic increase in fluctuation amplitudes, described in the foregoing section, motivates an investigation of changes in the resulting radial turbulent transport. 
It is commonly characterized by a radial particle diffusivity $D_r$ and heat conductivities $\chi_\mathrm{r}^\mathrm{e,i}$, which are critical input parameters in mean-field transport simulations \cite{wiesen2015,bufferand2013,feng2014,pan2022,rivals2024,feng2024,winters2024}. 
Since the radial transport in turbulence simulations is an output, we may compute \textit{effective} particle diffusivity $D_r$ and heat conductivity $\chi_r$ for our simulations in post. These we define as
\begin{align}
    D_r &= - \frac{\tilde{\Gamma}_r}{\langle \partial_r n \rangle} = - \frac{\langle n \mathbf{v}_E \cdot \mathbf{e}_r\rangle - \langle n \rangle \langle \mathbf{v}_E \cdot \mathbf{e}_r\rangle}{\langle \partial_r n \rangle} \,, \\
    \chi_{r,s} &= - \frac{\tilde{Q}_{r,s} - \frac{3}{2} \langle T_s \rangle \tilde{\Gamma}_r}{\langle n \rangle \langle \partial_r T_s \rangle} =  - \frac{3}{2} \frac{\langle n T_s \mathbf{v}_E \cdot \mathbf{e}_r \rangle   - \langle T_s \rangle\langle n \mathbf{v}_E \cdot \mathbf{e}_r \rangle}{\langle n \rangle \langle \partial_r T_s \rangle} \, ,
\end{align}
where $s \in \{\mathrm{e,i}\}$ and angular brackets $\langle \circ \rangle$ denote averages over time and toroidal angle.
Figure \ref{fig:diffusivities} shows the resulting $D_r$ and electron heat conductivity $\chi_{r,\mathrm{e}}$ for the No-XPR and High-XPR cases. Radial ion conductivity $\chi_{r,\mathrm{i}}$, while not shown, is qualitatively identical to $\chi_{r, \mathrm{e}}$.

We find that diffusivities generally increase in detachment when compared to the attached reference case. 
The effect is most pronounced in the low-field side scrape-off layer, where radial heat conductivity increases by more than a factor of 10.
In comparison, effective diffusivities at the XPR structure itself are moderate in the High-XPR case. Note that this may still result in strong radial flows due to the presence of steep gradients at the XPR (see Figures \ref{fig:lo_xpr_snap} and \ref{fig:hi_xpr_snap}), which will be further discussed in section \ref{section:analysis:exbflows}.

In all simulations, the effective diffusivities vary strongly in space and also become negative. This indicates that turbulence cannot be well described by local diffusive mixing. Instead, it exhibits convective pinches, non-diagonal and non-local transport, including streamers and blob-filaments \cite{zholobenko2023}. These become particularly pronounced in detached conditions.

\subsection{Radial electric field}

\begin{figure}
    \centering
    \includegraphics[width=0.35\linewidth,trim={0 9 0 5},clip]{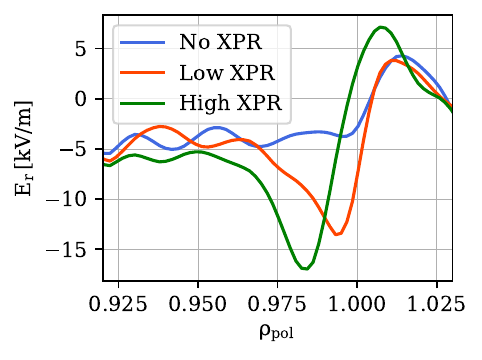}
    \includegraphics[width=0.30\linewidth,trim={0 2 0 5},clip]{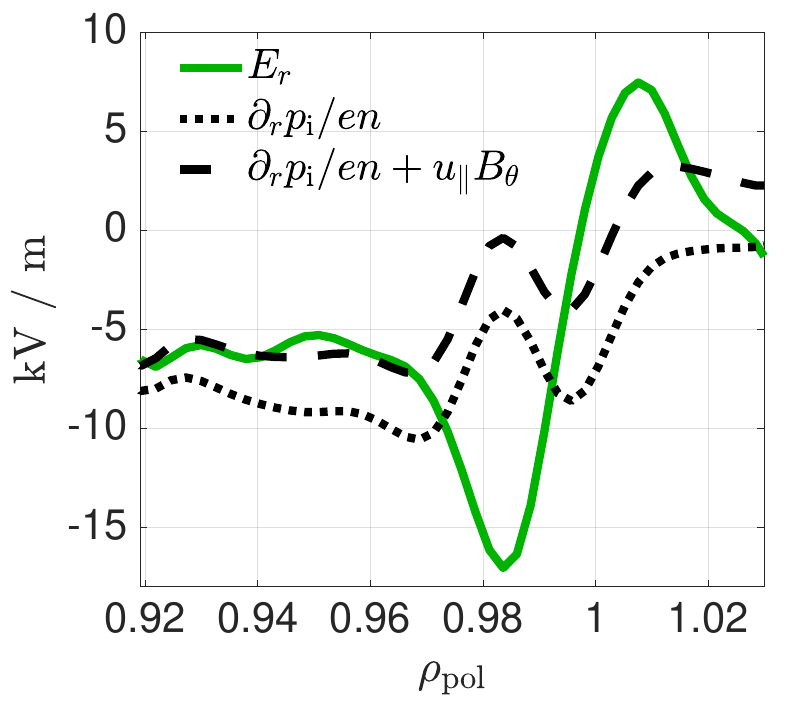}
    \caption{Left: time and toroidally averaged radial electric field $E_r$ at the OMP in the three simulations. Right: explicit contributions from the radial ion gradient (dotted line) and additionally toroidal rotation (dashed line) in the High-XPR case. The remaining difference to $E_r$ (green solid line) is due to poloidal rotation $u_\mathrm{\theta} B_\mathrm{tor}$, indicating turbulence-driven zonal flows in the field well at $\rho_\mathrm{pol} = 0.98$.}
    \label{fig:erad}
\end{figure}

In previous SOLPS-ITER transport simulations \cite{senichenkov2021b, pan2022}, the existence of an XPR was found to be related to an inward shift of the $E_r$ shear layer at the OMP.
Similar inward movement of $E_r$ was also observed in turbulence simulations of detached conditions using a half-size TCV setup \cite{mancini2023}.
We confirm these results, plotting in Figure \ref{fig:erad} the time and toroidally averaged $E_r$ at the OMP. Compared to either the No- or Low-XPR simulations, the $E_r$ well for the High-XPR case distinctly shifts inward by $0.5\,\mathrm{cm}$.
Additionally, let us briefly discuss its composition. In the confined region, the time-averaged radial electric field is determined by the average radial ion pressure gradient, toroidal rotation, and poloidal rotation \cite{zholobenko2021a, zholobenko2023},
\begin{equation}\label{eq:er}
    \langle E_r \rangle_{t, \varphi} \approx \langle \frac{\partial_r p_\mathrm{i}}{e n} \rangle_{t, \varphi} + \langle u_\parallel B_\theta \rangle_{t, \varphi} - \langle  u_\theta B_\mathrm{tor} \rangle_{t, \varphi} \, .
\end{equation}
Taking a more detailed look at the High-XPR simulation (right subplot of Figure \ref{fig:erad}), we find that $E_r$ in the core region up to $\rho_\mathrm{pol} = 0.96$ is well-explained by the former two contributions (dashed line).
However, they fail to reproduce the electric field well ($\rho_\mathrm{pol} = 0.98$) by themselves, and poloidal rotation $u_\theta B_\mathrm{tor}$ becomes significant. The latter, in turn, is the result of zonal flows driven by turbulence via the Reynolds stress \cite{diamond2005}. 
As they do not vanish in the time average, zonal flows play a key role in determining poloidal rotation in the edge. This phenomenon is consistent with prior turbulence simulations in attached conditions \cite{zholobenko2021a, zhang2024, zholobenko2024} and not exclusive to the detached XPR state.

\subsection{$E \times B$ flows near the X-point radiator}
\label{section:analysis:exbflows}
\begin{figure}
    \centering
    \includegraphics[width=0.99\linewidth,trim={0 5 0 5},clip]{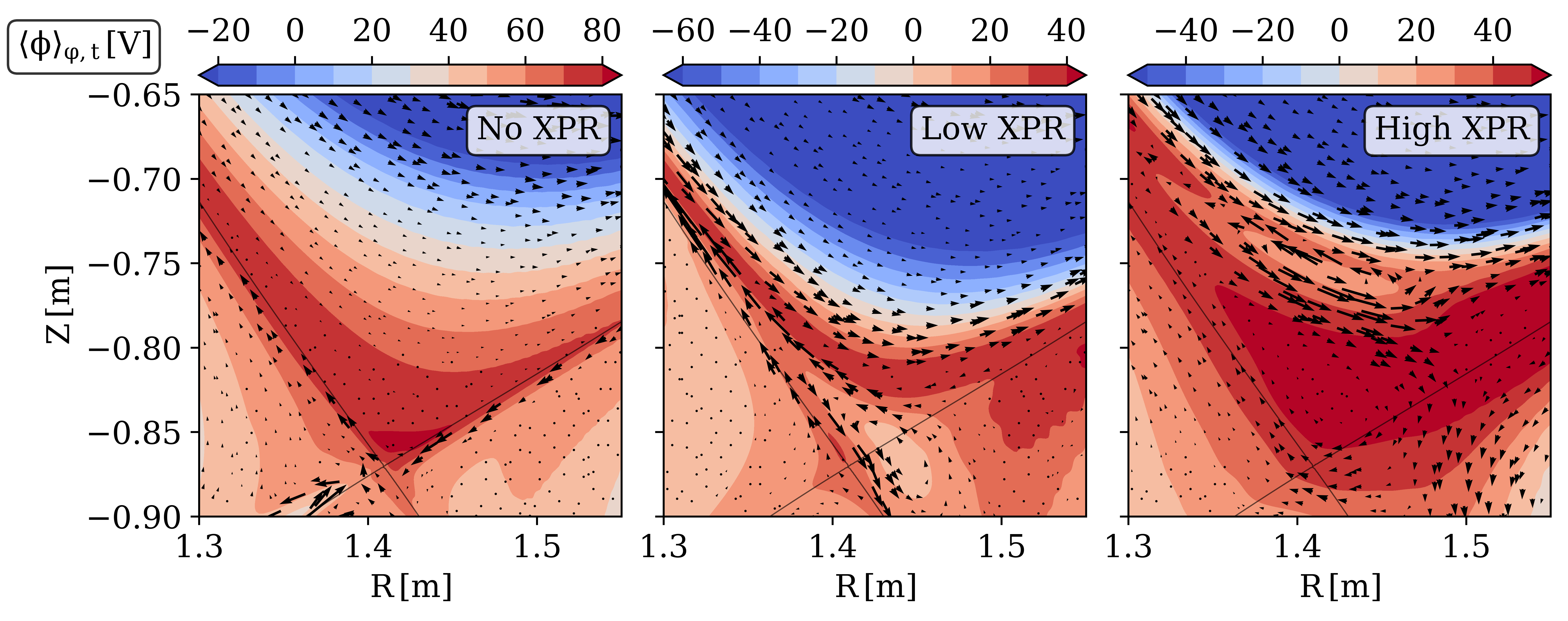}
    \caption{Magnified view of time and toroidally averaged electrostatic potential above the X-point. The averaged $E\times B$ particle flux $n \mathbf{v}_\mathrm{E}$ is overlaid with black arrows.}
    \label{fig:2d_potential}
\end{figure}

Finally, in Figure \ref{fig:2d_potential}, we show the averaged electrostatic potential $\langle \phi \rangle_{\varphi, t}$ near the X-point and overlay the associated $E\times B$ particle flux $\langle n \left(\mathbf{B} \times \nabla_\perp \phi \right) / B^2 \rangle_{\varphi, t}$. Note that the view is further magnified compared to the fluctuation amplitudes shown in Figure \ref{fig:2d_fluctuation}.
In the No-XPR case, the potential in the confined region mostly aligns with the flux surfaces, indicating regular poloidal rotation. 
This alignment is disrupted once the detachment front rises into the confined region, as first shown in the Low-XPR case, and becomes pronounced in the High-XPR case.
In the latter, the distinct poloidal structure becomes fully contained within the confined region, developing into a stationary $E\times B$ vortex cell featuring a counterclockwise flow pattern.
Its location overlaps with the dense center of the XPR, where density fluctuations are maximized (top row of Figure \ref{fig:2d_fluctuation}). 

\begin{figure}
    \centering
    \includegraphics[width=0.5\linewidth,trim={0 8 0 5},clip]{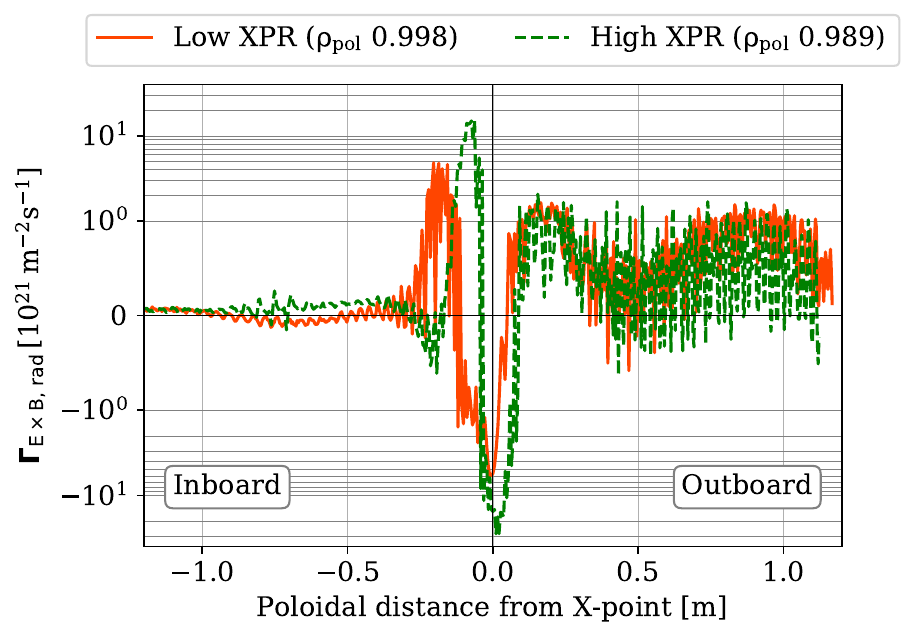}
    \caption{Radial particle flux through particular flux surfaces in the two simulations featuring an XPR\blue{, positive values indicate radially outward flow}. The inboard and outboard mid-planes are located at distances $\sim \pm 1\,\mathrm{m}$ relative to the X-point. Note that the $y$-axis first scales linearly between $[-10^0, 10^0]$ and thereafter scales logarithmically in both directions.}
    \label{fig:radflux}
\end{figure}

The averaged drift flows thus reveal that the XPR is associated with significant local transport across flux surfaces. 
We further highlight this effect in Figure \ref{fig:radflux}, plotting the radial $E\times B$ particle flux through a single flux surface (chosen to intersect with the XPR core, i.e.~$\rho_\mathrm{pol} = 0.998$ in the Low-XPR case, $\rho_\mathrm{pol} = 0.989$ in the High-XPR case).
The outboard side features typical turbulent radial transport outward (positive $\mathbf{\Gamma}_\mathrm{E\times B, rad}$) and suppression of radial transport on the inboard side \cite{zholobenko2023}.
The No-XPR case, while not shown here, yields much the same asymmetry of radial transport. 
Exclusively in the two XPR simulations, however, we also find significant radial flows near the X-point, which exceed regular outboard transport by up to an order of magnitude and correspond to the convective cells shown in Figure \ref{fig:2d_potential}. 
Forthcoming studies will focus more closely on the radial transport of particles (and energy) under such detached X-point radiating conditions.

\section{Conclusions}
\label{section:conclusion}

We have performed the first turbulence simulations of detached X-point radiator (XPR) conditions, providing insight into the turbulent nature of the XPR phenomenon.
Such simulations required several extensions to the GRILLIX plasma edge turbulence code. 
We included plasma cooling by impurity radiation and modified the treatment of neutral gas diffusion and parallel resistivity to improve the numerical stability of the code. 
The turbulence simulations discussed herein are based on the L-mode phase of a fully detached  ASDEX Upgrade discharge with an XPR.
Two such simulations are shown, with radiating fronts at $5\,\mathrm{cm}$ (referred to as ``Low-XPR'' case) and $12\,\mathrm{cm}$ (referred to as the ``High-XPR'' case) above the X-point.
The XPR structure observed in the simulations is consistent with previous analytical and SOLPS-ITER transport studies, which describe the XPR as having a mantle of ionization and impurity radiation surrounding a cold core of recombining plasma at the center. 
However, the turbulence simulations reveal a novel feature of the XPR: recombining cores are highly intermittent, and multiple may exist simultaneously along the front. The ionization and radiation mantles fluctuate accordingly, although the two-dimensional structures observed in previous SOLPS-ITER transport simulations are recovered when averaging over time and toroidal angle.
The Low- and High-XPR simulations, complemented by an attached reference simulation (``No-XPR'' case), are validated against experimental measurements at the outboard-midplane and divertor. 
Good agreement with the experimental data is observed, especially for the High-XPR case.
Furthermore, we compare the simulated radiation density in the divertor against bolometry measurements of the X-point radiator.
The High-XPR simulation again yields excellent agreement and reproduces the peaked radiation profile of the experiment.

The turbulence observed near the X-point is then examined in more detail.
\blue{The distributions of density and temperature fluctuations in the vicinity of the XPR become significantly broader, with density fluctuations exceeding 400\% of the background.} 
\red{Consequently, the turbulent transport is also enhanced, and is quantified by effective diffusivities. The transport also exhibits strong non-local and non-diffusive behavior, however.} 
\red{In agreement with previous transport studies, we find an inward shift of the radial electric field profile associated with the XPR state. The XPR also affects the electrostatic potential at the X-point, which develops a poloidally asymmetric structure. As a consequence, large radial flows in a vortex pattern are observed at the XPR, which exceed the turbulent radial transport at the outboard side by a factor of 10. We conclude that these effects warrant immediate further study, as they may have implications for the ELM suppression in H-mode XPR conditions.}


\section{Acknowledgements}

The authors gratefully acknowledge Kaiyu Zhang, Christoph Pitzal, Jan Pfennig and Barnab\'{a}s Csillag for discussions on data interpretation and visualization.
We further thank Bernd Kurzan and Ou Pan for assisting in experimental data acquisition. Finally, we thank Tom Body for valuable discussions on impurity radiation coefficients.
This work has been carried out within the framework of the EUROfusion Consortium, funded by the European Union via the Euratom Research and Training Programme (Grant Agreement No 101052200 – EUROfusion). 
Views and opinions expressed are those of the author(s) only and do not necessarily reflect those of the European Union or the European Commission. 
Neither the European Union nor the European Commission can be held responsible for them. 
The simulations shown in this work were performed on the national supercomputer HPE Apollo (Hawk) at the High Performance Computing Center Stuttgart (HLRS) under the grant number GRILLIX/44281, and the EUROfusion High-Performance Computer Marconi-Fusion (A3) under the project TSVV3.

\section{Declaration of generative AI and AI-assisted technologies in the writing process}
During the preparation of this work, the author(s) used DeepL Write
and ChatGPT to polish the language. After using this tool/service, the author(s) reviewed and edited the content as needed and take(s) full responsibility for the content of the publication.

\section{Appendix}
\label{appendix}

\subsection{Numerical treatment of parallel resistivity in Ohm's Law}
\label{parallel_resistivity}

As discussed in Section \ref{section:model}, treating parallel resistivity in Ohm's law implicitly was found to be important for affordable simulations in detached conditions. This is briefly detailed here. Highlighting only key terms, the equation may be abbreviated as
\begin{equation}\label{eq:ohm_simplified}
    \beta_0 \frac{\partial}{\partial t} A_{\|}
    + \mu \left( 
        \frac{\partial}{\partial t} + \mathbf{v}_E \cdot \nabla + v_\parallel \nabla_\parallel 
        \right) \frac{j_\parallel}{n}
    + \left( \frac{\eta_{\parallel 0}}{T_\mathrm{e}^{3/2}} \right) j_\parallel
    = \mathrm{RHS} \,,
\end{equation}
describing the time evolution of the parallel vector potential $A_\parallel$ and parallel current $j_\parallel$, which are coupled via Amp\'ere's Law $j_\parallel = - \nabla_\perp^2 A_\parallel$. The remaining variables denote plasma density $n$, electron temperature $T_\mathrm{e}$, $E \times B$ drift velocity $\mathbf{v}_E$, electron parallel velocity $v_\parallel$, and mass ratio $\mu = m_\mathrm{e} / m_\mathrm{i}$. The reference electron beta, $\beta_0=\mu_0 n_0T_\mathrm{e0}/B_0^2$, and normalized parallel resistivity, $\eta_{\parallel 0} = 0.51 \mu \nu_\mathrm{e0}$, with electron collision time $\nu_\mathrm{e0}$, are constant parameters due to the normalization of the equations. A detailed description of the normalization scheme and remaining right-hand side terms $\mathrm{RHS}$ can be found in Appendix A.~of \cite{zholobenko2024}.
For illustrative purposes, we demonstrate the implicit treatment of parallel resistivity with an implicit Euler scheme, while in practice, the 3rd-order Karniadakis scheme is used for time-stepping \cite{stegmeir2019}. The part of Ohm's law leading to the restriction is 
\begin{equation}
\beta_0 \frac{\partial}{\partial t} A_\parallel = \left( \frac{\eta_{\parallel 0}}{T_\mathrm{e}^{3/2}} \right) \nabla_\perp^2 A_\parallel \,,
\end{equation}
which is a diffusion equation on $A_\parallel$ with the CFL condition $\Delta t \leq \frac{1}{4} \Delta x_\perp^2 \frac{\beta_0}{\eta_{\parallel0}} T_\mathrm{e}^{3/2}$, where $\Delta x_\perp$ is the (normalized) perpendicular grid distance. The implicitly discretized Ohm's law then reads
\begin{equation}
    \left[ 
        \beta_0 
        - \left( \frac{\mu}{n} + \Delta t \frac{\eta_{\parallel 0}}{T_\mathrm{e}^{3/2}} \right)^{(t+\Delta t)} \nabla_\perp^2
    \right] A_\parallel^{(t+\Delta t)}
    = \left[ 
        \beta_0 A_\parallel 
        + \mu \frac{\nabla_\perp^2 A_\parallel}{n} 
        - \Delta t \, \mu \left( \mathbf{v}_E \cdot \nabla + v_\parallel \nabla_\parallel \right) \frac{j_\parallel}{n} 
        + \Delta t \, \mathrm{RHS}
    \right]^{(t)} \, .
\end{equation}
$n^{(t+\Delta t)}$ and $T_\mathrm{e}^{(t+\Delta t)}$ are obtained by evolving them in time before solving for $A_\parallel^{(t+\Delta t)}$. 
Therefore, a generalized elliptic equation in perpendicular planes has to be solved to obtain $A_\parallel^{(t+\Delta t)}$, for which an efficient solver with a geometric multigrid preconditioner is available in GRILLIX \cite{stegmeir2019}. The implementation has been verified with the Method of Manufactured Solutions \cite{roache2002}.


\section*{References}
\bibliographystyle{rackstyle}
\bibliography{bibliography}

\newcommand{\noop}[1]{}
\begin{thebibliography}{10}

\bibitem{viezzer2023}
E.~Viezzer, M.~Austin, M.~Bernert \emph{et~al.}
\newblock Prospects of core--edge integrated no-ELM and small-ELM scenarios for
  future fusion devices.
\newblock \emph{Nuclear Materials and Energy}, \textbf{34}, 101308 (2023).

\bibitem{zohm2013}
H.~Zohm, C.~Angioni, E.~Fable \emph{et~al.}
\newblock On the physics guidelines for a tokamak DEMO.
\newblock \emph{Nuclear Fusion}, \textbf{53}~(7), 073019 (2013).

\bibitem{pitts2019}
R.~A. Pitts, X.~Bonnin, F.~Escourbiac \emph{et~al.}
\newblock {Physics basis for the first ITER tungsten divertor}.
\newblock \emph{Nuclear Materials and Energy}, \textbf{20}, 100696 (2019).

\bibitem{loarte2007}
A.~Loarte, B.~Lipschultz, A.~S. Kukushkin \emph{et~al.}
\newblock Progress in the ITER Physics Basis - Chapter 4: Power and particle
  control.
\newblock \emph{Nuclear Fusion}, \textbf{47}~(6), S203 (2007).

\bibitem{wischmeier2015}
M.~Wischmeier, A.~U. Team \emph{et~al.}
\newblock High density operation for reactor-relevant power exhaust.
\newblock \emph{Journal of Nuclear Materials}, \textbf{463}, 22 (2015).

\bibitem{stangeby2000}
P.~C. Stangeby \emph{et~al.}
\newblock \emph{The Plasma Boundary of Magnetic Fusion Devices}, volume 224.
\newblock Institute of Physics Pub. Philadelphia, Pennsylvania (2000).

\bibitem{kuang2020}
A.~Kuang, S.~Ballinger, D.~Brunner \emph{et~al.}
\newblock Divertor heat flux challenge and mitigation in SPARC.
\newblock \emph{Journal of Plasma Physics}, \textbf{86}~(5), 865860505 (2020).

\bibitem{zohm2021}
H.~Zohm, F.~Militello, T.~Morgan \emph{et~al.}
\newblock The EU strategy for solving the DEMO exhaust problem.
\newblock \emph{Fusion Engineering and Design}, \textbf{166}, 112307 (2021).

\bibitem{krasheninnikov2017}
S.~Krasheninnikov and A.~Kukushkin.
\newblock Physics of ultimate detachment of a tokamak divertor plasma.
\newblock \emph{Journal of Plasma Physics}, \textbf{83}~(5), 155830501 (2017).

\bibitem{pshenov2017}
A.~Pshenov, A.~Kukushkin and S.~Krasheninnikov.
\newblock Energy balance in plasma detachment.
\newblock \emph{Nuclear Materials and Energy}, \textbf{12}, 948 (2017).

\bibitem{wiesen2015}
S.~Wiesen, D.~Reiter, V.~Kotov \emph{et~al.}
\newblock The new SOLPS-ITER code package.
\newblock \emph{Journal of Nuclear Materials}, \textbf{463}, 480 (2015).

\bibitem{bufferand2013}
H.~Bufferand, B.~Bensiali, J.~Bucalossi \emph{et~al.}
\newblock Near wall plasma simulation using penalization technique with the
  transport code SOLEDGE2D-EIRENE.
\newblock \emph{Journal of Nuclear Materials}, \textbf{438}, S445 (2013).

\bibitem{feng2014}
Y.~Feng, H.~Frerichs, M.~Kobayashi \emph{et~al.}
\newblock Recent improvements in the EMC3-Eirene code.
\newblock \emph{Contributions to Plasma Physics}, \textbf{54}~(4-6), 426
  (2014).

\bibitem{bucalossi2022}
J.~Bucalossi, J.~Achard, O.~Agullo \emph{et~al.}
\newblock Operating a full tungsten actively cooled tokamak: overview of WEST
  first phase of operation.
\newblock \emph{Nuclear Fusion}, \textbf{62}~(4), 042007 (2022).

\bibitem{stroth2022scenario}
U.~Stroth, D.~Aguiam, E.~Alessi \emph{et~al.}
\newblock Progress from ASDEX Upgrade experiments in preparing the physics
  basis of ITER operation and DEMO scenario development.
\newblock \emph{Nuclear Fusion}, \textbf{62}~(4), 042006 (2022).

\bibitem{wersal2015}
C.~Wersal and P.~Ricci.
\newblock A first-principles self-consistent model of plasma turbulence and
  kinetic neutral dynamics in the tokamak scrape-off layer.
\newblock \emph{Nuclear Fusion}, \textbf{55}~(12), 123014 (2015).

\bibitem{kvist2024}
K.~Kvist, A.~S. Thrys{\o}e, T.~Haugb{\o}lle \emph{et~al.}
\newblock A direct Monte Carlo approach for the modeling of neutrals at the
  plasma edge and its self-consistent coupling with the 2D fluid plasma edge
  turbulence model HESEL.
\newblock \emph{Physics of Plasmas}, \textbf{31}~(3) (2024).

\bibitem{quadri2024}
V.~Quadri, P.~Tamain, Y.~Marandet \emph{et~al.}
\newblock Self-organization of plasma edge turbulence in interaction with
  recycling neutrals.
\newblock \emph{Contributions to Plasma Physics}, \textbf{64}~(7-8), e202300146
  (2024).

\bibitem{eder2025}
K.~Eder, A.~Stegmeir, W.~Zholobenko \emph{et~al.}
\newblock Self-consistent plasma-neutrals fluid modeling of edge and scrape-off
  layer turbulence in diverted tokamaks.
\newblock \emph{Plasma Physics and Controlled Fusion}, \textbf{67}~(6), 065034
  (2025).

\bibitem{mancini2023}
D.~Mancini, P.~Ricci, N.~Vianello \emph{et~al.}
\newblock Self-consistent multi-component simulation of plasma turbulence and
  neutrals in detached conditions.
\newblock \emph{Nuclear Fusion}, \textbf{64}~(1), 016012 (2023).

\bibitem{kallenbach1995}
A.~Kallenbach, R.~Dux, V.~Mertens \emph{et~al.}
\newblock H mode discharges with feedback controlled radiative boundary in the
  ASDEX Upgrade tokamak.
\newblock \emph{Nuclear Fusion}, \textbf{35}~(10), 1231 (1995).

\bibitem{lowry1997}
C.~Lowry, D.~Campbell, S.~Davies \emph{et~al.}
\newblock Divertor configuration studies on JET.
\newblock \emph{Journal of Nuclear Materials}, \textbf{241}, 438 (1997).

\bibitem{pitts1999}
R.~Pitts, A.~Refke, B.~Duval \emph{et~al.}
\newblock Experimental investigation of the effects of neon injection in TCV.
\newblock \emph{Journal of Nuclear Materials}, \textbf{266}, 648 (1999).

\bibitem{park2023}
J.-S. Park, R.~Pitts, J.~Jang \emph{et~al.}
\newblock Bifurcation-like transition of divertor conditions induced by X-point
  radiation in KSTAR L-mode plasmas.
\newblock \emph{Nuclear Fusion}, \textbf{63}~(8), 086018 (2023).

\bibitem{goetz1996}
J.~Goetz, C.~Kurz, B.~LaBombard \emph{et~al.}
\newblock Comparison of detached and radiative divertor operation in Alcator
  C-Mod.
\newblock \emph{Physics of Plasmas}, \textbf{3}~(5), 1908 (1996).

\bibitem{reimold2015a}
F.~Reimold, M.~Wischmeier, M.~Bernert \emph{et~al.}
\newblock Divertor studies in nitrogen induced completely detached H-modes in
  full tungsten ASDEX Upgrade.
\newblock \emph{Nuclear Fusion}, \textbf{55}~(3), 033004 (2015).

\bibitem{gloeggler2019}
S.~Gl{\"o}ggler, M.~Wischmeier, E.~Fable \emph{et~al.}
\newblock Characterisation of highly radiating neon seeded plasmas in JET-ILW.
\newblock \emph{Nuclear Fusion}, \textbf{59}~(12), 126031 (2019).

\bibitem{bernert2021}
M.~Bernert, F.~Janky, B.~Sieglin \emph{et~al.}
\newblock X-point radiation, its control and an ELM suppressed radiating regime
  at the ASDEX Upgrade tokamak.
\newblock \emph{Nuclear Fusion}, \textbf{61}~(2), 024001 (2021).

\bibitem{lipschultz1984}
B.~Lipschultz, B.~LaBombard, E.~Marmar \emph{et~al.}
\newblock Marfe: an edge plasma phenomenon.
\newblock \emph{Nuclear Fusion}, \textbf{24}~(8), 977 (1984).

\bibitem{bernert2025}
M.~Bernert, T.~Bosman, T.~Lunt \emph{et~al.}
\newblock X-point radiation: From discovery to potential application in a
  future reactor.
\newblock \emph{Nuclear Materials and Energy}, \textbf{43}, 101916 (2025).

\bibitem{bernert2023}
M.~Bernert, S.~Wiesen, O.~F{\'e}vrier \emph{et~al.}
\newblock The X-Point radiating regime at ASDEX Upgrade and TCV.
\newblock \emph{Nuclear Materials and Energy}, \textbf{34}, 101376 (2023).

\bibitem{rivals2024}
N.~Rivals, N.~Fedorczak, P.~Tamain \emph{et~al.}
\newblock Experiments and SOLEDGE3X modeling of dissipative divertor and
  X-point Radiator regimes in WEST.
\newblock \emph{Nuclear Materials and Energy}, \textbf{40}, 101723 (2024).

\bibitem{bosman2024}
T.~Bosman, M.~Bernert, L.~Ceelen \emph{et~al.}
\newblock X-point radiator control and its dynamics in ASDEX Upgrade and JET
  deuterium–tritium discharges.
\newblock \emph{Nuclear Fusion}, \textbf{65}~(1), 016057 (2024).

\bibitem{reimerdes2024}
H.~Reimerdes, C.~Theiler, M.~Bernert \emph{et~al.}
\newblock Access to an ELM-suppressed X-point radiator regime in TCV snowflake
  minus configurations.
\newblock \emph{Nuclear Materials and Energy}, \textbf{41}, 101784 (2024).

\bibitem{senichenkov2021a}
I.~Y. Senichenkov, E.~Kaveeva, V.~Rozhansky \emph{et~al.}
\newblock Approaching the radiating X-point in SOLPS-ITER modeling of ASDEX
  Upgrade H-mode discharges.
\newblock \emph{Plasma Physics and Controlled Fusion}, \textbf{63}~(5), 055011
  (2021).

\bibitem{pan2022}
O.~Pan, M.~Bernert, T.~Lunt \emph{et~al.}
\newblock SOLPS-ITER simulations of an X-point radiator in the ASDEX Upgrade
  tokamak.
\newblock \emph{Nuclear Fusion}, \textbf{63}~(1), 016001 (2022).

\bibitem{sun2023}
G.~Sun.
\newblock SOLPS-ITER simulation of an X-point radiator in TCV.
\newblock arXiv preprint: \url{https://arxiv.org/abs/2311.07295} (2023).

\bibitem{poletaeva2024}
A.~Poletaeva, V.~Rozhansky, E.~Kaveeva \emph{et~al.}
\newblock First SOLPS-ITER modelling of an X-point radiator in ITER.
\newblock \emph{Nuclear Fusion}, \textbf{64}~(12), 126038 (2024).

\bibitem{feng2024}
Y.~Feng, V.~Winters, D.~Zhang \emph{et~al.}
\newblock Conditions and benefits of X-point radiation for the island divertor.
\newblock \emph{Nuclear Fusion}, \textbf{64}~(8), 086027 (2024).

\bibitem{winters2024}
V.~Winters, F.~Reimold, Y.~Feng \emph{et~al.}
\newblock First experimental confirmation of island SOL geometry effects in a
  high radiation regime on W7-X.
\newblock \emph{Nuclear Fusion}, \textbf{64}~(12), 126047 (2024).

\bibitem{kobayashi2013}
M.~Kobayashi, S.~Masuzaki, I.~Yamada \emph{et~al.}
\newblock Control of 3D edge radiation structure with resonant magnetic
  perturbation fields applied to the stochastic layer and stabilization of
  radiative divertor plasma in LHD.
\newblock \emph{Nuclear Fusion}, \textbf{53}~(9), 093032 (2013).

\bibitem{faitsch2021}
M.~Faitsch, T.~Eich, G.~Harrer \emph{et~al.}
\newblock Broadening of the power fall-off length in a high density, high
  confinement H-mode regime in ASDEX Upgrade.
\newblock \emph{Nuclear Materials and Energy}, \textbf{26}, 100890 (2021).

\bibitem{greenwald1999}
M.~Greenwald, R.~Boivin, P.~Bonoli \emph{et~al.}
\newblock Characterization of enhanced D$\alpha$ high-confinement modes in
  Alcator C-Mod.
\newblock \emph{Physics of Plasmas}, \textbf{6}~(5), 1943 (1999).

\bibitem{lunt2023}
T.~Lunt, M.~Bernert, D.~Brida \emph{et~al.}
\newblock Compact radiative divertor experiments at ASDEX Upgrade and their
  consequences for a reactor.
\newblock \emph{Physical Review Letters}, \textbf{130}~(14), 145102 (2023).

\bibitem{stegmeir2019}
A.~Stegmeir, A.~Ross, T.~Body \emph{et~al.}
\newblock Global turbulence simulations of the tokamak edge region with
  GRILLIX.
\newblock \emph{Physics of Plasmas}, \textbf{26}~(5) (2019).

\bibitem{stegmeir2016}
A.~Stegmeir, D.~Coster, O.~Maj \emph{et~al.}
\newblock The field line map approach for simulations of magnetically confined
  plasmas.
\newblock \emph{Computer Physics Communications}, \textbf{198}, 139 (2016).

\bibitem{stegmeir2023}
A.~Stegmeir, T.~Body and W.~Zholobenko.
\newblock Analysis of locally-aligned and non-aligned discretisation schemes
  for reactor-scale tokamak edge turbulence simulations.
\newblock \emph{Computer Physics Communications}, \textbf{290}, 108801 (2023).

\bibitem{body2019}
T.~Body, A.~Stegmeir, W.~Zholobenko \emph{et~al.}
\newblock Treatment of advanced divertor configurations in the flux-coordinate
  independent turbulence code GRILLIX.
\newblock \emph{Contributions to Plasma Physics}, \textbf{60}~(5-6), e201900139
  (2019).

\bibitem{stegmeir2025}
A.~Stegmeir, M.~Finkbeiner, C.~Pitzal \emph{et~al.}
\newblock GRILLIX as unified fluid turbulence code for tokamaks and
  stellarators.
\newblock \emph{Available at SSRN 5104975} (2024).

\bibitem{zhang2024}
K.~Zhang, W.~Zholobenko, A.~Stegmeir \emph{et~al.}
\newblock Magnetic flutter effect on validated edge turbulence simulations.
\newblock \emph{Nuclear Fusion}, \textbf{64}~(3), 036016 (2024).

\bibitem{zholobenko2024}
W.~Zholobenko, K.~Zhang, A.~Stegmeir \emph{et~al.}
\newblock Tokamak edge-SOL turbulence in H-mode conditions simulated with a
  global, electromagnetic, transcollisional drift-fluid model.
\newblock \emph{Nuclear Fusion}, \textbf{64}~(10), 106066 (2024).

\bibitem{zholobenko2021b}
W.~Zholobenko, A.~Stegmeir, M.~Griener \emph{et~al.}
\newblock The role of neutral gas in validated global edge turbulence
  simulations.
\newblock \emph{Nuclear Fusion}, \textbf{61}~(11), 116015 (2021).

\bibitem{braginskii1965}
S.~Braginskii.
\newblock Transport processes in a plasma.
\newblock \emph{Reviews of Plasma Physics}, \textbf{1}, 205 (1965).

\bibitem{zeiler1997}
A.~Zeiler, J.~F. Drake and B.~Rogers.
\newblock Nonlinear reduced Braginskii equations with ion thermal dynamics in
  toroidal plasma.
\newblock \emph{Physics of Plasmas}, \textbf{4}~(6), 2134 (1997).

\bibitem{horsten2017}
N.~Horsten, G.~Samaey and M.~Baelmans.
\newblock Development and assessment of 2D fluid neutral models that include
  atomic databases and a microscopic reflection model.
\newblock \emph{Nuclear Fusion}, \textbf{57}~(11), 116043 (2017).

\bibitem{uytven2020}
W.~Van~Uytven, M.~Blommaert, W.~Dekeyser \emph{et~al.}
\newblock Implementation of a separate fluid-neutral energy equation in
  SOLPS-ITER and its impact on the validity range of advanced fluid-neutral
  models.
\newblock \emph{Contributions to Plasma Physics}, \textbf{60}~(5-6), e201900147
  (2020).

\bibitem{stotler2017}
D.~Stotler, J.~Lang, C.~Chang \emph{et~al.}
\newblock Neutral recycling effects on ITG turbulence.
\newblock \emph{Nuclear Fusion}, \textbf{57}~(8), 086028 (2017).

\bibitem{zhang2020}
Y.~Zhang and S.~I. Krasheninnikov.
\newblock Effect of neutrals on the anomalous edge plasma transport.
\newblock \emph{Plasma Physics and Controlled Fusion}, \textbf{62}~(11), 115018
  (2020).

\bibitem{stroth2022xpr}
U.~Stroth, M.~Bernert, D.~Brida \emph{et~al.}
\newblock Model for access and stability of the X-point radiator and the
  threshold for marfes in tokamak plasmas.
\newblock \emph{Nuclear Fusion}, \textbf{62}~(7), 076008 (2022).

\bibitem{stroth2025}
U.~Stroth, M.~Bernert, T.~Lunt \emph{et~al.}
\newblock Model for the X-point radiator height and its energetic coupling to
  the plasma edge.
\newblock \emph{Plasma Physics and Controlled Fusion}, \textbf{67}~(3), 035001
  (2025).

\bibitem{adas}
Atomic Data and Analysis Structure.
\newblock \url{https://open.adas.ac.uk/}.

\bibitem{makarov2024}
S.~O. Makarov.
\newblock \emph{Development and Implementation of a Generalized Multi-Ion
  Transport Model in Plasma Edge Fluid Codes}.
\newblock Ph.D. thesis, Technische Universit{\"a}t M{\"u}nchen (2024).

\bibitem{fischer2010}
R.~Fischer, C.~Fuchs, B.~Kurzan \emph{et~al.}
\newblock Integrated Data Analysis of Profile Diagnostics at ASDEX Upgrade.
\newblock \emph{Fusion Science and Technology}, \textbf{58}~(2), 675 (2010).

\bibitem{fan2019}
D.~Fan, Y.~Marandet, P.~Tamain \emph{et~al.}
\newblock Effect of turbulent fluctuations on neutral particles transport with
  the TOKAM3X-EIRENE turbulence code.
\newblock \emph{Nuclear Materials and Energy}, \textbf{18}, 105 (2019).

\bibitem{zholobenko2023}
W.~Zholobenko, J.~Pfennig, A.~Stegmeir \emph{et~al.}
\newblock Filamentary transport in global edge-SOL simulations of ASDEX
  Upgrade.
\newblock \emph{Nuclear Materials and Energy}, \textbf{34}, 101351 (2023).

\bibitem{bernert2014axuv}
M.~Bernert, T.~Eich, A.~Burckhart \emph{et~al.}
\newblock Application of AXUV diode detectors at ASDEX Upgrade.
\newblock \emph{Review of scientific Instruments}, \textbf{85}~(3) (2014).

\bibitem{senichenkov2021b}
I.~Senichenkov, E.~Kaveeva, V.~Rozhansky \emph{et~al.}
\newblock Features of radial electric field in impurity-seeded, detached plasma
  in a tokamak.
\newblock \emph{Physics of Plasmas}, \textbf{28}~(6) (2021).

\bibitem{zholobenko2021a}
W.~Zholobenko, T.~Body, P.~Manz \emph{et~al.}
\newblock Electric field and turbulence in global Braginskii simulations across
  the ASDEX Upgrade edge and scrape-off layer.
\newblock \emph{Plasma Physics and Controlled Fusion}, \textbf{63}~(3), 034001
  (2021).

\bibitem{diamond2005}
P.~H. Diamond, S.~Itoh, K.~Itoh \emph{et~al.}
\newblock Zonal flows in plasma—a review.
\newblock \emph{Plasma Physics and Controlled Fusion}, \textbf{47}~(5), R35
  (2005).

\bibitem{roache2002}
P.~J. Roache.
\newblock Code Verification by the Method of Manufactured Solutions.
\newblock \emph{Journal of Fluids Engineering}, \textbf{124}~(1), 4 (2002).

\end{thebibliography}

\end{document}